\documentclass[linenumbers,trackchanges,twocolumn]{raa}
\usepackage{longtable}    
\usepackage{rotating}
\usepackage{pdflscape}      
\usepackage{booktabs}       
\usepackage{graphicx,times}
\usepackage{natbib}
\usepackage{array}  
\usepackage{amssymb,amsmath}
\bibpunct{(}{)}{;}{a}{}{,}

\newenvironment{acknowledgments}{\section*{Acknowledgments}}{}

\usepackage[pagebackref=true]{hyperref}
\usepackage[header,title,page,titletoc]{appendix}

\usepackage{orcidlink}

\begin{document}

\title{An extended catalogue of Herbig-Haro objects}
 \volnopage{ {\bf 20XX} Vol.\ {\bf X} No. {\bf XX}, 000--000}
   \setcounter{page}{1}

\author{
    Zhi-Yu Chen \inst{1} \and Chao-Jian Wu \orcidlink{0000-0003-3514-6619} \inst{1,*}\footnotetext{$*$Corresponding Authors, these authors contributed equally to this work.}
}

\institute{CAS Key Laboratory of Optical Astronomy, National Astronomical Observatories, Chinese Academy of Sciences, Beijing 100101, China; {\it chjwu@bao.ac.cn}\\
             \vs\no
   {\small Received 202x month day; accepted 202x month day}}

\abstract{We present an extended catalogue of 1193 Herbig–Haro (HH) objects, comprising 477 isolated HH objects and 716 HH knots, compiled through a meticulous review of the literature available through mid-2025. The catalogue provides comprehensive data for each entry, including celestial coordinates, distances, knot separations, exciting sources (with spectral types where available), object characteristics, and bibliographic references.
We also perform a preliminary statistical analysis of key parameters such as distance, exciting source properties, morphology, and excitation state. 
By combining our extended catalogue with the two earlier HH object catalogues published by Hippel et al. in 1988 and Reipurth in 2000, astronomers can access comprehensive information on all known HH objects, thereby facilitating research on star formation and stellar outflows.
}
 \authorrunning{Z.-Y. Chen \& C.-J. Wu}            
 \titlerunning{An extended HHOs Catalogue}  
 \maketitle

\section{Introduction} \label{sec:intro}

Herbig-Haro (HH) objects are small emission nebulae associated with star-forming regions, typically found within dense molecular clouds and observed in the vicinity of young stars. First identified in the 1950s by George Herbig \citep{herbig1951spectra} and Guillermo Haro \citep{haro1952herbig, haro1953halpha}, HH objects are produced when energetic outflows from young stellar objects (YSOs) interact with the surrounding interstellar medium (ISM). These interactions generate shock waves that produce visible emission in the form of luminous knots, detectable across multiple wavelengths, particularly in the optical and infrared, as well as at longer wavelengths.

Since 2000, one of the most notable advances in HH object research has been the refinement of observational techniques. The advent of high-resolution imaging and spectroscopy, along with space-based observatories such as the Hubble Space Telescope (HST) \citep{livio2003world}, has significantly enhanced our ability to resolve and examine the detailed structures of HH objects. These instruments have enabled accurate measurements of jet velocities, shock temperatures, and ionization states within HH outflows. In particular, infrared observations from the Spitzer \citep{werner2004spitzer} and Herschel space telescopes \citep{pilbratt2010herschel} have revolutionized the detection of HH objects in deeply embedded star-forming regions, which are heavily obscured at optical wavelengths by dust extinction.

Ground-based observatories have likewise played a pivotal role in advancing HH object studies. The development and application of adaptive optics (AO) and integral field spectroscopy (IFS) have allowed astronomers to study these objects in unprecedented detail, yielding high-resolution imaging and detailed kinematic maps that reveal their complex morphological structures.
Research on HH objects extends well beyond optical observations. For example, using high-resolution data from the Atacama Large Millimeter/submillimeter Array (ALMA) \citep{fomalont20152014} on six dusty pillars and globules in the Carina Nebula, \cite{2023ApJ...958..193C} detected molecular outflows along with their associated dusty envelopes and disks. These observations further revealed multiple millimeter continuum sources in the HH~900 and HH~1004 regions, indicating several distinct outflows and thereby offering a broader analysis that builds upon previous findings in this area.
Multi-epoch studies of the HH~211 protostellar jet with the Submillimeter Array (SMA) support a scenario in which its knots represent internal shocks caused by periodic variations in the jet ejection velocity, likely driven by the orbital motion of the jet source \citep{2016ApJ...816...32J}.

\cite{1988A&AS...74..431V} published a catalogue of HH objects identified prior to 1987 in 1988. Subsequently, \cite{2000yCat.5104....0R} released a new catalogue of HH objects in 2000, which has been periodically updated. These two compilations represent the most comprehensive catalogues of HH objects available to date. However, no complete catalogue has been established to include HH objects discovered since 2000, leaving a significant gap in the reference materials available to researchers. This absence hinders the systematic collection of statistical data on HH objects and complicates large-sample studies. To address this gap, we have conducted a systematic review of all literature pertaining to the discovery and study of HH objects published between 2000 and the present and have compiled the findings into an expanded catalogue. Combining our catalogue with the two previously mentioned compilations will enable astronomers to perform statistical analyses and conduct more detailed studies of all known HH objects. This integrated resource plays a crucial role in advancing the understanding of star formation, outflows, and mass-loss processes.

In this study, we present an updated compilation and analysis of HH objects observed since 2000, incorporating a total of 659 individual HH objects (477 HH objects lacking knots and 182 HH objects exhibiting knots). Counted by the number of knots, the total can reach up to 1100. Our compilation includes detailed information on celestial coordinates, distances, spectral classifications, and associated young stellar sources. Through a systematic examination of these properties, we aim to provide a comprehensive overview of the spatial distribution and physical characteristics of HH objects. Additionally, this study assesses how technological and observational advances over the past two decades have improved our understanding of these dynamically evolving astrophysical phenomena.

\section{Identifying HH Objects}

    HH objects are small-scale, shock-excited nebulae associated with regions containing young stellar objects or protostars. Historically, the first recorded HH object is Burnham's Nebula, discovered through visual observation in 1890 and later cataloged as HH 255 \citep{burnham1890note}. The class of HH objects was established following the spectroscopic surveys by Herbig and Haro \citep{magakian2016searches}.

    Observationally, HH objects are detected and defined by their optical emission lines, which include [O\,\textsc{i}], [O\,\textsc{ii}], [O\,\textsc{iii}], [N\,\textsc{ii}], and [S\,\textsc{ii}]-particularly the [S\,\textsc{ii}] doublet at 6717 \AA and 6731 \AA-as well as the Balmer lines H$\alpha$ and H$\beta$. These lines are characteristic of shock-excited gas and serve as the primary spectroscopic diagnostic for identifying HH objects.

    Although some HH objects are detected in near-infrared emission lines, outflows observed solely in the near-infrared are not classified as HH objects but rather as Molecular Hydrogen Emission‑line Objects (MHOs). This distinction ensures that the HH catalogue remains mainly focused on optically bright, shock-excited nebulae.

    Morphologically, HH objects are typically classified into three types: bow shocks, jets, and knots, which help to identify the nature of the outflow source \citep{reipurth1991herbig}. High-resolution observations reveal complex structures within HH outflows, such as bow shock wings, Mach disks, and cooling zones. The morphology of HH objects provides essential clues for determining the outflow source and its interaction with the ambient medium \citep{wang2000progress}.

    According to \cite{2000yCat.5104....0R} and our catalogue, most outflows have velocities in the range of 50-300 km $\textsc{s}^{-1}$, consistent with shock velocities expected in these young stellar environments.

    The cataloguing of HH objects follows established conventions to ensure clarity. Outflows with clear spatial structure (greater than approximately 1 arcsecond) are assigned individual names, such as HH 1, HH 2, and HH 3. For internal labelling, distinct knots or features within a single flow are designated with capital letters-for example, HH 1A, HH 1B, HH 1C \citep{magakian2016searches, herbig1974draft}.

    However, with the arrival of the JWST era, the definition of HH objects has evolved, as these new facilities reveal shocked emission from [Fe\,\textsc{ii}] and H$_2$ that is often obscured in the optical. This shift likely calls for a redefinition of the classical HH object and requires that the methods for searching for and studying HH objects also adapt and advance accordingly.

\section{Catalogue} \label{sec:cat}

\subsection{Catalogue compilation} \label{subsec:catcom}
This study employed a literature review methodology, analyzing more than 
one hundred papers published between 2000 and March 2025 concerning newly identified HH objects. 
Data on celestial coordinates, distances, knots separation, morphology and exciting sources were 
extracted from these publications. Additional parameters, including excitation 
types and line intensity ratios, were also recorded when available. 
The exciting sources of many HH objects remain unidentified. Most of the known exciting sources are detected in the infrared band and lack optical information, likely due to extinction from the surrounding interstellar medium. Consequently, their stellar types cannot be determined. In such cases, we only provide the stellar ID in the table for reference by future researchers.

The extended catalogue of HH objects compiled in this paper contains 1193 entries (including 477 isolated HH objects and 716 HH object knots). 
Given the considerable length of the table, the full catalogue has been placed in the appendix at the end of the article.

To maintain completeness, the catalogue also includes references to MHOs associated with HH objects, thereby providing a more comprehensive view of outflows from young stellar objects across different wavelength.

Because the full catalogue is too long, Table \ref{tab:subcat} presents information only for a portion of HH objects, illustrating the overall structure of the catalogue. Designations, coordinates, distances, angular distance between knots, characteristic and references are all included in our HH objects catalogue. The full catalogue can be seen in appendix \ref{sec:maintable}. 

Table \ref{tab:subcat} and Appendix \ref{sec:maintable} are organized as follows:

\textit{Column 1, 2}: The name of the HH object and MHO, where the number indicates the designated serial number of the HH object and its corresponding MHO designations, if applicable. Letters following the number (e.g., A, B, C) denote that the same HH object may have distinct parts, which are also referred to as knots.
		
\textit{Columns 3, 4}: Right ascension and declination in equinox J2000 coordinates. The quoted coordinate accuracy reflects the uncertainty reported by the referenced authors. Some HH objects have differing coordinates across literature (e.g., HH 502); in such cases, we adopt the coordinates from the latest publication.

\textit{Columns 5}: Date of coordinate measurement.

\textit{Column 6}: distance to the observer, given in parsecs (pc). All distances are compiled from the references. 

\textit{Column 7}: the angular separation between HH knots. The HH objects may have many knots. 
Column 7 shows the angular separation, in arcsecs, between different knots of the same HH 
object. An asterisk (*) denotes the separation between the first knot and the exciting source, 
while all other values represent the separation from the previous knot listed in the table.

\textit{Column 8}: the column 8 describes the characteristics of HH objects and HH object knots, including descriptions of their morphology, size, brightness, and emission line intensity, etc.

\textit{Column 9}: the bibliographic references attributed to the first 
discovery or confirmation of each HH object.
\clearpage
\begin{landscape}
\centering
\footnotesize  
\setlength{\tabcolsep}{3pt} 
\setlength{\LTleft}{\fill}    
\setlength{\LTright}{\fill}   

\begin{longtable}{@{}llllcclll@{}} 

\caption{Extended catalogue of HH Objects}\\
\label{tab:subcat}\\
\toprule
Name & MHO & RA & DEC & Date & Dist & Angular separation between knots & Characteristic & References \\
(HH) & & (J2000) & (J2000) & & (pc) & (arcsec) & & \\
\midrule
\endfirsthead

\multicolumn{9}{l}{{\itshape Continued}} \\
\toprule
Name & MHO & RA & DEC & Date & Dist & Angular separation between knots & Characteristic & References \\
(HH) & & (J2000) & (J2000) & & (pc) & (arcsec) & & \\
\midrule
\endhead

\midrule
\multicolumn{9}{r}{{\itshape Continued on next page}} \\
\endfoot

\bottomrule
\endlastfoot

76c           &-     &15:00:35.2      &-63:01:26   & 2020 Oct         &700   &-   &linear                               &\citep{rector2020herbig}                       \\
76c2          &-     &15:00:31.7      &-63:01:39   & 2025 Mar         &700   &27.13    &-                                   &\citep{rector2025herbig}                       \\
76d           &-     &15:00:38.5      &-63:02:20   & 2020 Oct         &700   &61.81    &-                                   &\citep{rector2020herbig}                       \\
76e           &-     &15:00:37.6      &-63:03:11   & 2020 Oct         &700   &51.37    &-                                   &\citep{rector2020herbig}                       \\
76f           &-     &15:00:38.6      &-63:03:51   & 2020 Oct         &700   &40.57    &-                                   &\citep{rector2020herbig}                       \\
76g           &-     &15:00:39.0      &-63:04:42   & 2020 Oct         &700   &51.07    &-                                   &\citep{rector2020herbig}                       \\
76h           &-     &15:00:39.5      &-63:05:06   & 2020 Oct         &700   &24.24    &-                                   &\citep{rector2020herbig}                       \\
76i           &-     &15:00:33.4      &-62:58:26   & 2025 Mar         &700   &-        &-                                   &\citep{rector2025herbig}                       \\
77b           &-     &15:00:43.1      &-63:07:26   & 2025 Mar         &700   &-        &-                                   &\citep{rector2025herbig}                       \\
119D          &2404  &19:37:05.82     &07:34:07.2  & 2007 Nov         &250   &-        &H$\alpha$ bow shock                  &\citep{gaalfalk2007herbig,davis2010general}    \\
119E          &-     &19:37:09.19     &07:34:06.5  & 2007 Nov         &250   &50.11    &Leading feature bright in R band     &\citep{gaalfalk2007herbig}                     \\
119F          &-     &19:37:09.39     &07:33:46.0  & 2007 Nov         &250   &20.71    &Bow shock                            &\citep{gaalfalk2007herbig,davis2010general}    \\
119G          &2403  &19:37:04.75     &07:33:38.3  & 2007 Nov         &250   &69.42    &-                                   &\citep{gaalfalk2007herbig}                     \\
119H          &-     &19:37:10.22     &07:33:56.8  & 2007 Nov         &250   &83.41    &More extended at 4.5 $\mu$m          &\citep{gaalfalk2007herbig,davis2010general}    \\
119I          &2406  &19:36:45.41     &07:34:40.2  & 2007 Nov         &250   &371.45   &Part of a WNW flow                   &\citep{gaalfalk2007herbig}                     \\
182A          &327   &05:54:09.8      &02:37:09    & 2005 Jul         &460   &-        &Patch                               &\citep{wang2005new, davis2010general}         \\
182B          &-     &05:54:11.2      &02:37:15    & 2005 Jul         &460   &21.82    &Faint patch                         &\citep{wang2005new}                            \\
380 F         &-     &20:58:55.1      &+52:34:56   & 2010 Mar         &800   &-        &-                                    &\citep{magakian2010wide} \\
380 C         &-     &20:59:08.2      &+52:36:32   & 2010 Mar         &800   &153.1    &-                                    &\citep{magakian2010wide} \\
380 D         &-     &20:59:09.1      &+52:36:14   & 2010 Mar         &800   &19.8     &-                                    &\citep{magakian2010wide} \\
380 E         &910 C &20:59:15.0      &+52:35:06   & 2010 Mar         &800   &86.7     &-                                    &\citep{magakian2010wide} \\
381 jet       &-     &20:58:21.2      &+52:29:32   & 2010 Mar         &800   &-        &-                                    &\citep{magakian2010wide} \\
381 C         &-     &20:58:21.9      &+52:28:54   &2010 Mar         &800   &38.5     &-                                    &\citep{magakian2010wide} \\

\end{longtable}
\end{landscape}

\subsection{Kinematic information of HH objects}
    The kinematic properties of HH objects, such as proper motion, radial velocity, and angular position, are crucial for analyzing their motion characteristics and understanding the underlying outflow dynamics. A comprehensive assessment of these parameters provides essential insights into the excitation mechanisms, collimation, and evolution of the outflows. Accordingly, we present a summary table that compiles the key kinematic information for the HH objects discussed here.
    

    In particular, accurate radial velocities and proper motions offer a direct window into the physical processes at work, going far beyond what imaging alone can reveal. Proper motions trace the tangential motion of individual knots, allowing us to reconstruct the jet's spatial structure and to identify velocity asymmetries that may reflect variations in ejection history or interaction with the ambient medium. Radial velocities, on the other hand, provide the line of sight component of the motion, enabling a full three dimensional reconstruction of the outflow kinematics when combined with proper motions. Together, these measurements make it possible to estimate dynamical timescales, link outflow features to their driving sources, and constrain key physical parameters such as shock velocities and mass ejection rates.

    To facilitate such analyses, we present the table in Appendix \ref{sec:tab_vel} that compiles the key kinematic information for the HH objects discussed in this work. Appendix \ref{sec:tab_vel} contains six columns, listing respectively the HH object name, proper motion (in mas $\mathrm{yr^{-1}}$), position angle (P.A., in degrees), shock velocity (in km $\mathrm{s^{-1}}$), radial velocity (RV, in km $\mathrm{s^{-1}}$), and the corresponding references. Where multiple measurements exist in the literature, we include the most recent or most accurate values.

\section{Analysis}\label{sec:Analysis}
	
	\subsection{Space Distribution}
	
\begin{figure*}
        \centering
        \includegraphics[width=\textwidth]{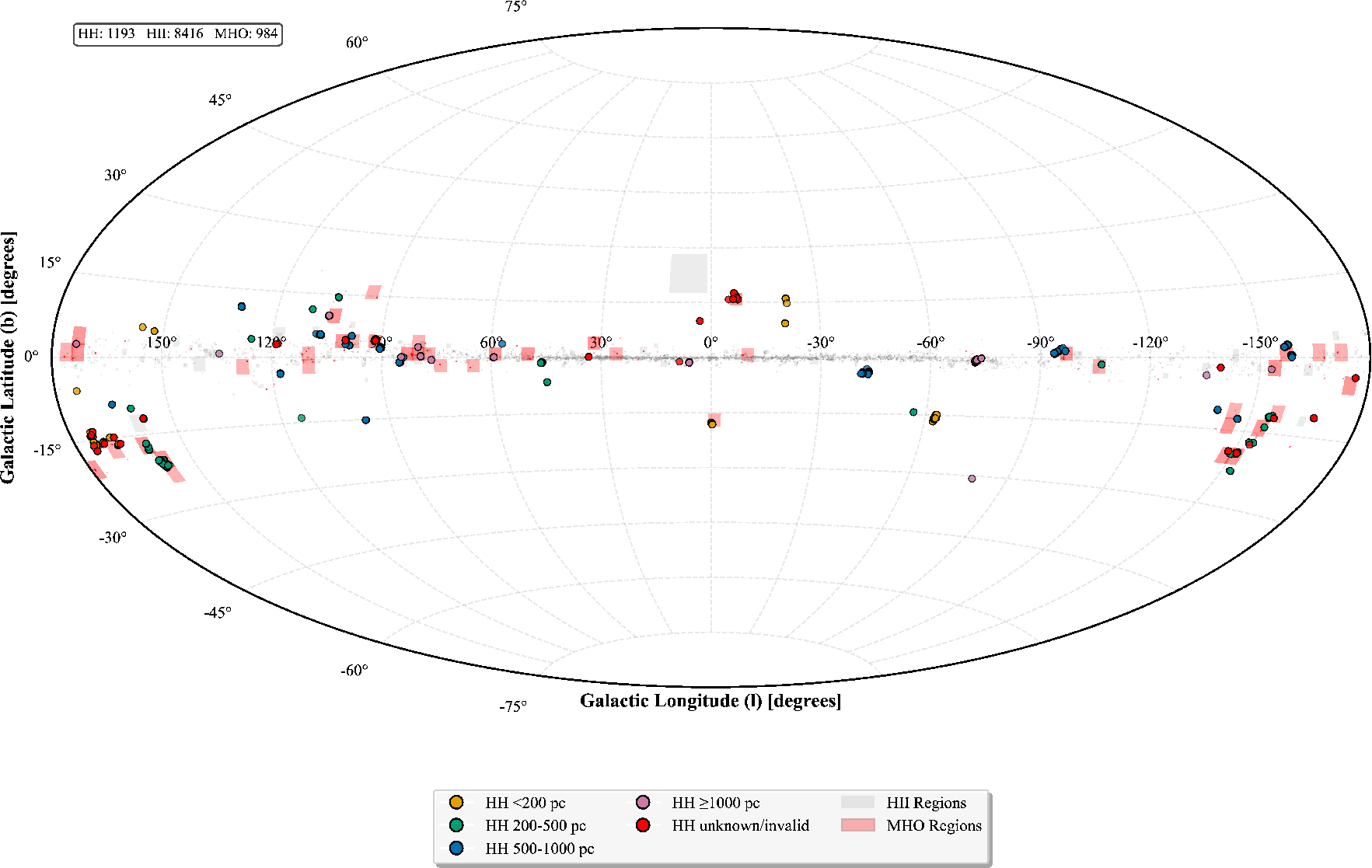}
        \caption{Spatial distribution of HH objects in Galactic coordinates. 
        The distribution shows that most known HH objects are located at high 
        Galactic latitudes ($|b|>5^\circ$). This is because the complex ISM in the 
        Galactic plane hinders their detection. The circles are color-coded by 
        distance, and the gray rectangles indicate H\,\textsc{ii} regions.}
        \label{fig:distr}
        \end{figure*}

	Based on the collected coordinates of the HH objects, we have plotted the spatial distribution of the HH objects, MHO (pink squares) and H\,\textsc{ii} regions (gray squares) on the celestial sphere (Figure \ref{fig:distr}).
		
    Dense nebulae in the cores of H\,\textsc{ii} regions outshine faint HH shocks, making them difficult 
    to detect. Cloud compression at ionization fronts triggers star 
    formation, which produces collimated jets. As a result, over 80\% of the cataloged HH 
    objects are located in these peripheral zones.

The spatial and distance distribution of HH objects is presented in Figure \ref{fig:distr}. The majority are found at Galactic latitudes between $-30^\circ$ and $15^\circ$, supporting a good correlation with active star-forming regions and the ISM, while the pink squares in Figure \ref{fig:distr} indicate the positions of MHO objects, revealing a strong spatial correlation between HH objects and MHO objects.
This distribution is notably asymmetric, with a higher concentration of objects detected south of the Galactic plane compared to the north. In the figure, the color of each point indicates its distance: orange (0-200 pc), green (200-500 pc), blue (500-1000 pc), purple ($>$1000 pc), with red marking objects of unknown distance. 
Analysis reveals that 45.22\% (298 HH objects) of our sample lie within $\pm 5^\circ$ of the Galactic plane, while an additional 51.59\% (340 HH objects) are located within $\pm 10^\circ$. This indicates that more than half of the sample resides beyond the immediate vicinity ($\pm 5^\circ$) of the Galactic plane.
The fact that the majority of detected HH objects reside outside known H\,\textsc{ii} regions suggests that the complex environment close to the Galactic plane may hinder their detection. Consequently, focusing searches north of the plane, where the observed density is currently lower, could prove a more fruitful strategy for discovering new HH objects.

\begin{figure*}
        \centering
        \includegraphics[width=0.75\linewidth]{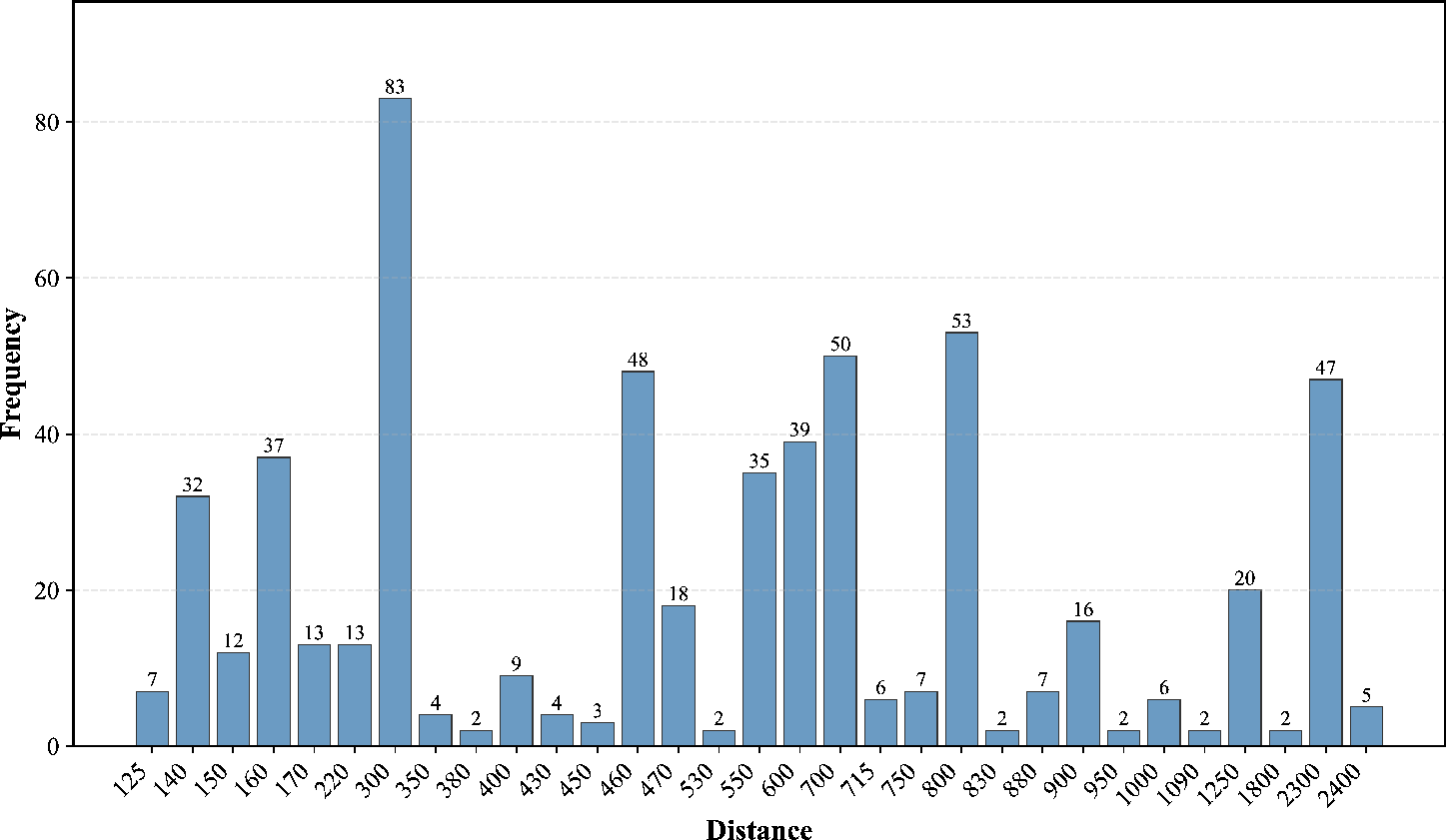}
        \caption{Distance distribution of HH objects in the extended catalogue. The distances of HH objects span a wide range, from 120 pc to 2500 pc. This distribution pattern may be related to the regions where HH objects are observed. Overall, no clear trend or regularity is evident in their distance distribution.}
        \label{fig:dist}
        \end{figure*}

For some HH objects, distance estimates are available directly from the literature. For some others, the distance can be inferred through their associated exciting sources. The histogram in Figure \ref{fig:dist} displays the distance distribution of HH objects. Analysis reveals significant clustering at distances of about 300 pc, 460 pc, 700 pc and 2300pc, corresponding to the Perseus, Orion, Circinus and Carina molecular cloud complexes, respectively. The abundance of HH objects in these regions provides direct evidence of ongoing active star formation within their embedded clusters, with the Orion Nebula Cluster being particularly prominent. Central to this process is the presence of high-density molecular clouds in these complexes, which supply the essential material primarily molecular hydrogen ($\mathrm{H_2}$), CO, and silicate dust for both protostellar accretion and the subsequent generation of HH jets. Furthermore, one extragalactic HH object (HH 1177) has been identified in the Large Magellanic Cloud (LMC). It lies at a distance of approximately 50,000 pc and has an estimated physical extent of about 10 pc \citep{mcleod2018parsec}. This is the only extragalactic HH object identified to date.

	\subsection{Exciting sources}

    The majority of HH objects in the extended catalogue lack an identified exciting source. Among them, 252 objects have at least one candidate exciting source identified. Of these, 8 objects are associated with multiple possible exciting sources, while the remaining 244 have a single confirmed exciting source. Details of these 252 HH objects are provided in Appendix \ref{sec:tab_host}.

    By cross-matching these exciting sources with Gaia data, we obtained spectral types for a subset of them, which have been listed in Appendix \ref{sec:tab_host}. Based on the available spectral classifications, six of the exciting sources are early-type stars (B/A type). Among them, LkH$\alpha$ 233 has excited seven HH objects (HH 808-814), and LkH$\alpha$ 234, with a spectral type of B5, has excited eight HH objects (HH 815-822). Another 19 exciting sources belong to FGK-type stars, of which AS353 (a K5-type star) has excited five HH objects (HH 1187-1191). The remaining 14 objects are associated with M dwarf stars. From these statistics, it can be observed that low- to intermediate-mass stars are more likely to excite HH objects, while massive stars tend to excite multiple HH object knots.

	\subsection{Excitation}
		
    The excitation states of HH objects, which are categorized as low, medium, or high excitation, were originally defined by \cite{raga1996compilation} based on diagnostic intensity ratios involving the emission lines of [O\,\textsc{iii}], H$\alpha$, H$\beta$, and [S\,\textsc{ii}].  We have applied this same classification criterion to determine the excitation states of the HH objects listed in Table \ref{tab:exci}. 

    As shown in Table \ref{tab:exci}, the majority of HH objects with known excitation types are classified as high excitation.  It is noteworthy that even different knots with the same exciting source can exhibit distinct classifications; for example, in HH 488, knot A is of high excitation, while knot B is classified as low excitation. 

    These excitation categories are derived from the relative intensities of specific emission lines.  The details regarding spectral line intensities and the associated statistical analysis are provided below.

    \clearpage
    \begin{landscape}
        \begin{longtable}{lrrrrl}
        \caption{The line intensity ratio of HH objects}\label{tab:exci} \\
        \hline
        Object &
        $[S_{\textsc{ii}}] \lambda6717/\lambda6731$ & 
        $[S_{\textsc{ii}}]\lambda\lambda(6717+6731)/H\alpha$ & 
        $[N_{\textsc{ii}}]\lambda\lambda(6548+6583)/H\alpha$ & 
        $[O_{\textsc{i}}]\lambda\lambda(6300+6363)/H\alpha$ &
        Excitation \\
        \hline
        \endfirsthead
        \caption[]{The line intensity ratio of HH objects (Continued)} \\
        \hline
        Object &
        $[S_{\textsc{ii}}] \lambda6717/\lambda6731$ & 
        $[S_{\textsc{ii}}]\lambda\lambda(6717+6731)/H\alpha$ & 
        $[N_{\textsc{ii}}]\lambda\lambda(6548+6583)/H\alpha$ & 
        $[O_{\textsc{i}}]\lambda\lambda(6300+6363)/H\alpha$ &
        Excitation \\
        \hline
        \endhead
        \hline
        \multicolumn{6}{r}{Continued on next page} \\
        \endfoot
        \hline
        \endlastfoot
        119D & -- & $1.749\pm0.077$ & -- & -- & -- \\
        119E & -- & $1.749\pm0.077$ & -- & -- & -- \\
        119F & -- & $1.749\pm0.077$ & -- & -- & -- \\
        119G & -- & $1.749\pm0.077$ & -- & -- & -- \\
        119I & -- & $1.749\pm0.077$ & -- & -- & -- \\
        182A & $1.19 \pm 0.13$ & $2.0 \pm 0.18$ & -- & -- & low \\
        434A & -- & 1.12 & 0.41 & 0.40 & -- \\
        435 & -- & 1.41 & 0.50 & 0.64 & -- \\
        436 & -- & 0.61 & 0.22 & 0.37 & -- \\
        439A & $1.16 \pm 0.04$ & $0.67 \pm 0.04$ & -- & -- & high \\
        450 & -- & 1.11 & -- & -- & -- \\
        450X & -- & $1.0\sim2.0$ & -- & -- & -- \\
        487A & -- & -- & -- & -- & high \\
        488A & -- & -- & -- & -- & high \\
        488B & -- & -- & -- & -- & low \\
        502N1 & -- & -- & -- & -- & high \\
        506 & -- & -- & -- & -- & low \\
        508 & -- & -- & -- & -- & high \\
        526 & -- & -- & -- & -- & high \\
        528 & -- & -- & -- & -- & high \\
        540A & -- & -- & -- & -- & low \\
        567C & -- & -- & -- & 0.34 & intermediate \\
        572A & 1.21 & 0.65 & 0.65 & -- & high \\
        575B & 1.46 & 0.48 & 0.66 & -- & high \\
        575C1 & 1.27 & 0.81 & 1.03 & -- & high \\
        586 & -- & -- & -- & 0.32 & intermediate \\
        587A & -- & -- & -- & 0.23 & -- \\
        588 center & -- & -- & -- & 0.23 & low \\
        588 NE1 & -- & -- & -- & 0.05 & high \\
        588 NE2 & -- & -- & -- & 0.08 & -- \\
        588 SW1 & -- & -- & -- & 0.10 & high \\
        588 SW2 & -- & -- & -- & 0.12 & -- \\
        589A & -- & -- & -- & 0.40 & -- \\
        589B & -- & -- & -- & 0.13 & -- \\
        666D & -- & $<0.04$ & -- & -- & -- \\
        666A & -- & 0.10 & -- & -- & -- \\
        666E & -- & 0.19 & -- & -- & -- \\
        666M & -- & 0.13 & -- & -- & -- \\
        666O & -- & 0.15 & -- & -- & -- \\
        666N & -- & 0.16 & -- & -- & -- \\
        666I & -- & 0.26 & -- & -- & -- \\
        666C & -- & 0.20 & -- & -- & -- \\
        688A & -- & -- & -- & -- & high \\
        704A & -- & -- & -- & -- & low \\
        727 & -- & -- & -- & -- & low \\
        866 & -- & -- & -- & -- & low \\
        892 & -- & -- & -- & -- & high \\
        965 & -- & -- & -- & -- & high \\
        979 NW1 & -- & -- & -- & -- & high \\
        992A & -- & -- & -- & -- & low \\
        1042 & -- & -- & -- & -- & high \\
        1043 & -- & -- & -- & -- & high \\
        1165 NW1 & 1.28 & 2.11 & 0.50 & -- & -- \\
        1165 NW2 & 1.34 & 2.31 & 0.36 & -- & -- \\
        1165 NW3 & 1.32 & 0.38 & -- & -- & -- \\
        1165 NW4 & 1.23 & 1.17 & -- & -- & -- \\
        1165 NW5 & 1.21 & 2.71 & 0.42 & -- & -- \\
        1165 SE1 & 1.42 & 1.20 & 1.46 & -- & -- \\
        \end{longtable}
        \end{landscape}

        \clearpage
    The [S\,\textsc{ii}] $\lambda\lambda$6716,6731 doublet ratio serves as a primary diagnostic for electron density. The observed values, clustering around 1.20, indicate that the sampled regions reside in a relatively uniform density regime, typically corresponding to densities near the low-density limit for these transitions.

    The ratio [S\,\textsc{ii}]/H$\alpha$ is a tracer of shock-dominated ionization. With the exception of HH 666, which exhibits a deviation suggestive of alternative ionization mechanisms, all other HH objects in our sample display ratios consistent with shock excitation. This alignment is expected given their high-velocity jet kinematics, which are conducive to strong shock formation.

    The [N\,\textsc{ii}] $\lambda\lambda$6548,6583/H$\alpha$ ratio, serves as a critical diagnostic tool in astrophysics, is primarily sensitive to the ionization parameter and metallicity of the emission interstellar gas. It can help distinguish between different excitation mechanisms: elevated ratios are often associated with harder radiation fields, such as those from active galactic nuclei (AGN) or shock excitation, whereas lower ratios are typical of photoionization by massive stars in H\,\textsc{ii} regions.
    In our sample, the [N\,\textsc{ii}] $\lambda\lambda$6548,6583/H$\alpha$ ratio is low. This is related to the excitation characteristics of HH objects: they are excited by low-velocity shocks, and the gas temperature and ionization degree in such shocks are insufficient to produce strong [N\,\textsc{ii}] emission.

    Finally, the [O\,\textsc{i}] $\lambda\lambda$ 6300,6363/H$\alpha$ ratio is sensitive to metallicity and excitation conditions. For the majority of targets, the derived values imply metallicities close to solar. However, notable outliers such as HH 434, HH 435, and HH 436 exhibit significantly enhanced ratios. This may indicate super-solar metallicities within these specific flows, potentially due to localized enrichment from processed stellar material or to shock-induced enhancement of the [O\,\textsc{i}] lines themselves. The aforementioned emission line ratios have all been dereddened.

\begin{figure}
        \centering
        \includegraphics[width=0.9\linewidth]{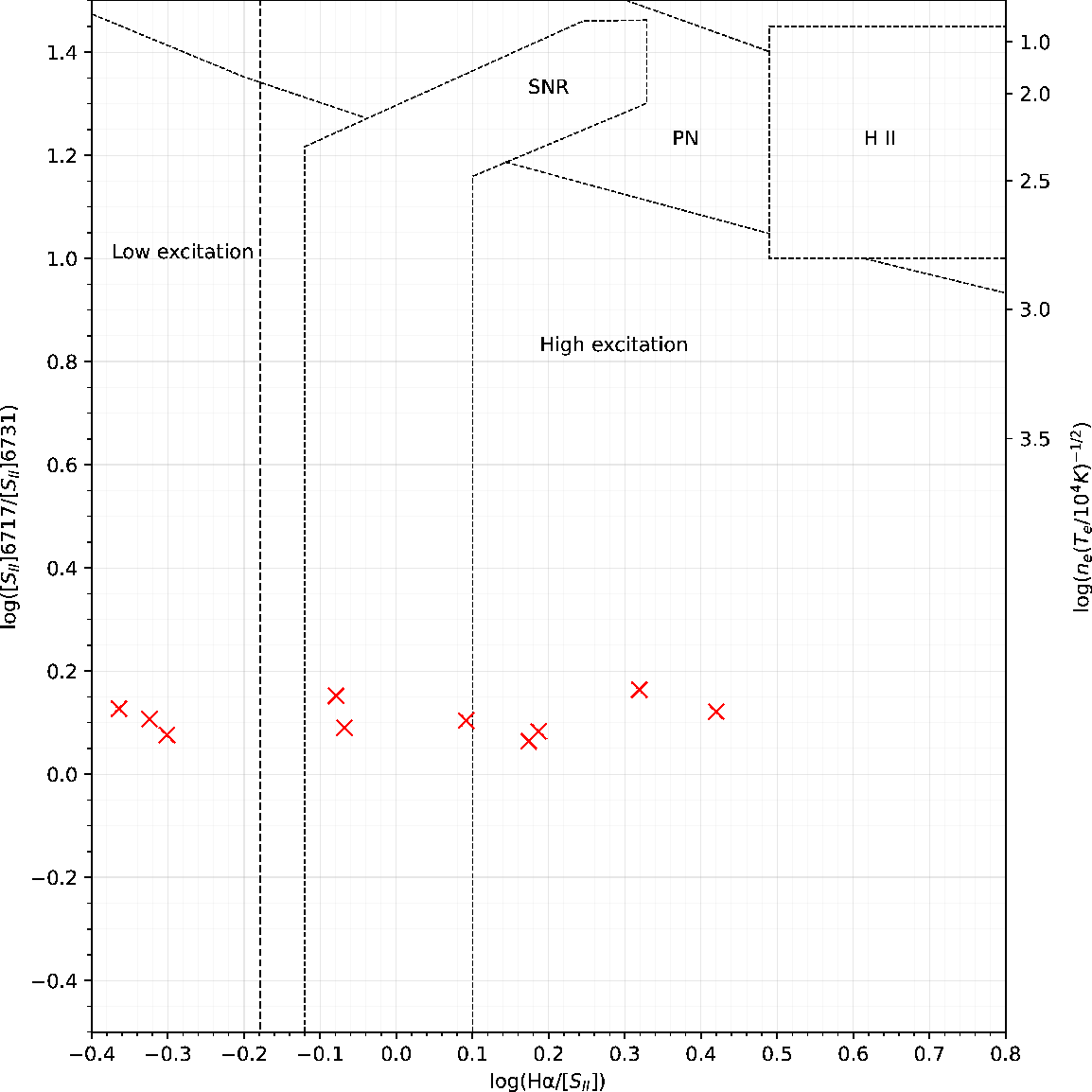}
        \caption{Excitation vs density diagnostic diagram \citep{lopez2009nature} for several HH objects in the extended catalogue.}
        \label{fig:diag1}
\end{figure}

\begin{figure}
        \centering
        \includegraphics[width=0.9\linewidth]{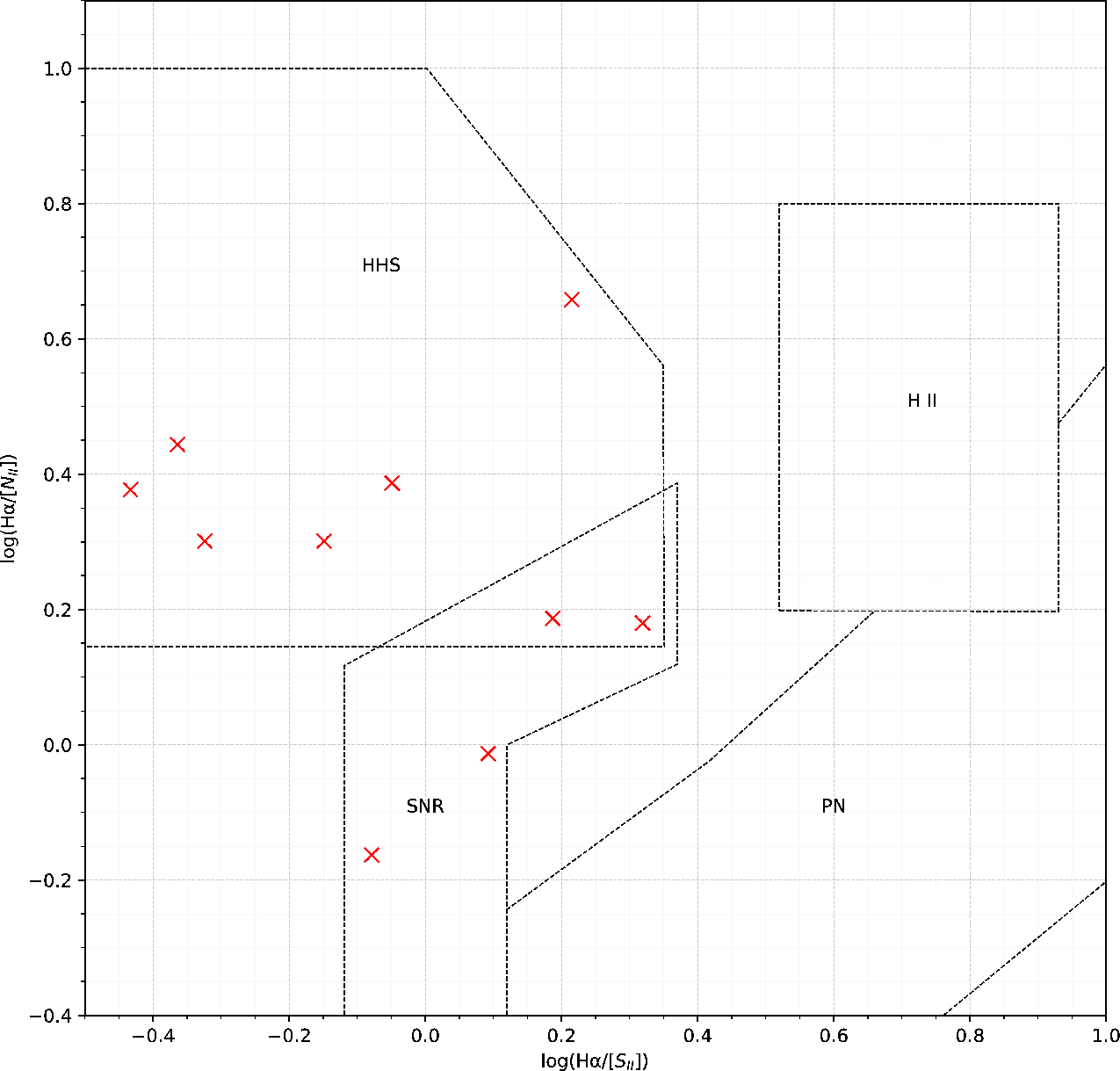}
        \caption{H$\alpha$/[S\,\textsc{ii}] vs H$\alpha$/[N\,\textsc{ii}] diagnostic diagram \citep{lopez2009nature} for distinguishing HH objects, SNRs, PNe and H\,\textsc{ii} regions.}
        \label{fig:diag2}
\end{figure}

    Based on the emission line ratios from Table \ref{tab:exci}, we plotted the diagnostic diagrams \citep{1981ASSL...91...95C, raga1996compilation}for H\,\textsc{ii} regions, planetary nebulae (PNe), and Supernova remnants (SNR). Figure \ref{fig:diag1} presents the [S\,\textsc{ii}] diagnostic diagram with log(H$\alpha$/[S\,\textsc{ii}]) on the abscissa and log([S\,\textsc{ii}]$\lambda$6717/[S\,\textsc{ii}]$\lambda$6731) on the ordinate. In this diagram, the plotted HH objects (primarily HH 182, HH 439, HH 572, HH 575, and HH 1165) exhibit a notable consistency in their [S\,\textsc{ii}]$\lambda$$\lambda$6717,6731 ratios. This clustering indicates that the ambient electron densities in their respective environments are relatively similar. Figure \ref{fig:diag2} shows a second diagnostic diagram with log(H$\alpha$/[S\,\textsc{ii}]) on the abscissa and log(H$\alpha$/[N\,\textsc{ii}]) on the ordinate, which includes demarcated zones for H\,\textsc{ii} regions, PNe, SNRs, and a specific zone for HH objects. 
    In Figure \ref{fig:diag2} all but two objects (HH575 C1 and HH1165 SE1) fall squarely within the defined HH object zone, reinforcing their distinct excitation properties consistent with shock physics.

		
	\subsection{Morphology}	 

    Of the 659 distinct HH objects analyzed, approximately 75\% (477 HH objects) show no detectable knots. When counting all individual knots, the total number of features increases to 1193. This implies that an HH object which does contain knots typically harbors an average of three to four such features. 
    Appendix \ref{sec:maintable} provides key geometric and morphological properties of the sampled HH objects. Column 7 lists the angular separation between adjacent knots within the same HH object, expressed in arcsec. Entries marked with an asterisk (*) specifically indicate the distance from a knot to its exciting source. 

    Column 8 describes the overall shape or pattern formed by these knots.
    Among the collected samples, most knot-bearing HH objects exhibit linear, bipolar, S-shaped, or C-shaped patterns in their knot distribution. Among the total sample, 44 HH objects or knots exhibit a linear morphology, accounting for approximately 3.69\% of the total. There are 108 HH objects or knots with S-shaped or C-shaped structures, representing about 9.05\% of the sample. Additionally, 152 objects or knots display bipolar (including 1 quadrupolar) morphologies, constituting roughly 12.74\% of the total. 
    It is noteworthy that HH 488A displays a quadrupolar morphology.

    Additionally, we analyzed the knot separations in HH objects driven by massive sources. While theoretical models predict larger knot separations for massive-star-driven jets, our analysis shows that objects excited by spectral type A sources do not present significantly greater separations compared to those driven by lower-mass sources. In contrast, HH objects associated with spectral type B stars exhibit markedly larger knot separations, substantially exceeding those found in other categories.
	
    \subsection{Near-infrared emission lines}

    Although HH outflows are primarily detected in optical emission lines such as H$\alpha$, [S\,\textsc{ii}], [O\,\textsc{i}], [O\,\textsc{iii}], [N\,\textsc{ii}], [Ca\,\textsc{ii}], and [C\,\textsc{i}], some have also been observed in near-infrared emission lines, including $\mathrm{H_2}$ and [Fe\,\textsc{ii}]. In most cases, near-infrared emission can be used to trace the outflow sources.

    In the Carina Nebula, 39 HH outflows have been detected in $\mathrm{H_2}$ emission, and 28 in [Fe\,\textsc{ii}] emission \citep{reiter2013hst}. 
    Some HH outflows have been detected in $\mathrm{H_2}$ emission, including: HH 727 \citep{hodapp2004disk}, 871 \citep{phelps2005parsec}, 382 E–G, 630 A, 631 F \citep{magakian2010wide}, 634 B, 634 C, 634 D, 635 A, 670 \citep{movsessian2003optical}, 777 \citep{reipurth2003blowout}, 493 \citep{walawender2004deep}, 576, 577, 908 B, 911, 924 \citep{bally2006deep}, 517, 528 \citep{bally2000disks}, 572, 575 \citep{wang2003herbig}, 469 \citep{aspinHerbig+-HaroFlowsCompact2000}, 625 \citep{o2003high}, 590, 593, 595 \citep{ogura2002halpha}, 182 \citep{wang2005new}, 1226, 1228, 1231 \citep{movsessian2024new}, 733 \citep{wang2004optical}, 240, 241 \citep{bally2009outflows}, 655, 656, 652, 647, 636, 637, 643 \citep{armond2011star}.

    Additionally, several HH outflows have been detected in [Fe\,\textsc{ii}] emission, such as: 574, 574X \citep{reipurth2004deep}, 1042, 1043 \citep{zhang2014herbig}, 1158 \citep{riaz2015hh}, 666 M, 666 O, 901, 902, 1066 \citep{reiter2013hst}, 567 C \citep{ogura2002halpha}, 903, 1006, 1005, 1010, 1014, 1160, 1163, 1164, 1167 \citep{reiter2017proper}, 900 \citep{reiterDisentanglingOutflowProtostars2015}, 1004, 1005, 1006, 1007, 1010, 1014, 1015, c-3, 1159, 1161, 1162, c-10, 1156 \citep{reiterFeIiJets2016}. 

    Because the near-infrared band is less affected by extinction and can probe deeper into the interstellar medium, searching for and studying outflows in the near-infrared will become a trend. In contrast, optical studies will gradually be superseded.
	

\section{Conclusion} \label{sec:conclusion}

    We present a comprehensive catalogue of 659 Herbig-Haro objects compiled from literature through mid-2025, significantly updating previous compilations. This catalogue serves as both a comparative reference for star formation studies and an indexed resource for existing observational data. Our analysis reveals that 45.22\% (298 HH objects) are concentrated within $|b| \leq 5^\circ$ of the Galactic plane, with 34.60\% of sources clustered at four characteristic distances (300\,pc, 460\,pc, 700\,pc and 2300\,pc). These distances correspond respectively to the Perseus, Orion, Circinus, and Carina nebular regions.
    The spectral types of the exciting sources have been compiled from the literature. 
    The subsequent statistical analysis reveals that low- to intermediate-mass stars are more likely to excite HH objects, while massive stars tend to excite multiple HH object knots.
    In our extended catalogue, 182 HH objects exhibit knots, with each HH object averaging 3-4 knots. It is worth noting that HH objects associated with Type B stars exhibit markedly larger knot separations. 

    The definition of HH objects has evolved with the advent of near- and mid-infrared facilities, where shocked emission from [Fe\,\textsc{ii}] and H$_2$ often serves as the primary tracer. However, as a summary work on HH objects studied primarily in the optical band, this catalogue will still serve as a useful reference for future research on outflows and shocks. 

\begin{acknowledgments}
This work is supported by the National Key R\&D Program of China 
2021YFA1600401, 2021YFA1600400, the National Natural Science
Foundation of China (NSFC) grant Nos. 12090041, 12090040, 12273052, 
Nos. 11733006, U1931109, 12003043, and the science research
grants from the China Manned Space Project (Nos. CMS-
CSST-2025-A14, CMS-CSST-2021-A04).

We would like to thank Miaomiao Zhang and Zhiwei Chen for helpful
suggestions.
\end{acknowledgments}
\appendix
\section{The full extended catalogue of HH objects}
\label{sec:maintable}
\begin{landscape}
\centering
\footnotesize  
\setlength{\tabcolsep}{3pt} 
\setlength{\LTleft}{\fill}    
\setlength{\LTright}{\fill}   

\begin{longtable}{@{}llllcclll@{}} 

\caption{Extended catalogue of HH Objects}\\
\label{tab:cat}\\
\toprule
Name & MHO & RA & DEC & Date & Dist & Angular separation between knots & Characteristic & References \\
(HH) & & (J2000) & (J2000) & & (pc) & (arcsec) & & \\
\midrule
\endfirsthead

\multicolumn{9}{l}{{\itshape Continued}} \\
\toprule
Name & MHO & RA & DEC & Date & Dist & Angular separation between knots & Characteristic & References \\
(HH) & & (J2000) & (J2000) & & (pc) & (arcsec) & & \\
\midrule
\endhead

\midrule
\multicolumn{9}{r}{{\itshape Continued on next page}} \\
\endfoot

\bottomrule
\endlastfoot
76c           &-     &15:00:35.2      &-63:01:26   & 2020 Oct         &700   &-   &linear                                                                                               &\citep{rector2020herbig}                       \\
76c2          &-     &15:00:31.7      &-63:01:39   & 2025 Mar         &700   &27.13    &-                                                                                               &\citep{rector2025herbig}                       \\ 
76d           &-     &15:00:38.5      &-63:02:20   & 2020 Oct         &700   &61.81    &-                                                                                               &\citep{rector2020herbig}                       \\ 
76e           &-     &15:00:37.6      &-63:03:11   & 2020 Oct         &700   &51.37    &-                                                                                               &\citep{rector2020herbig}                       \\ 
76f           &-     &15:00:38.6      &-63:03:51   & 2020 Oct         &700   &40.57    &-                                                                                               &\citep{rector2020herbig}                       \\ 
76g           &-     &15:00:39.0      &-63:04:42   & 2020 Oct         &700   &51.07    &-                                                                                               &\citep{rector2020herbig}                       \\ 
76h           &-     &15:00:39.5      &-63:05:06   & 2020 Oct         &700   &24.24    &-                                                                                               &\citep{rector2020herbig}                       \\ 
76i           &-     &15:00:33.4      &-62:58:26   & 2025 Mar         &700   &-        &-                                                                                               &\citep{rector2025herbig}                       \\
\end{longtable}
\end{landscape}
\section{The full Kinematic information catalogue of HH objects}
\label{sec:tab_vel}
\begin{landscape}
\footnotesize
\begin{longtable}{@{}llllll@{}}
\caption{Kinematic information of HH objects}\label{tab:vel_cat}\\
\hline
HHO & Proper motion & P.A. & Shock Velocity & Radial Velocity & Reference \\
 & (mas $\mathrm{yr^{-1}}$) & (degree) & (km $\mathrm{s^{-1}}$) & (km $\mathrm{s^{-1}}$) & \\
\hline
\endfirsthead
\caption[]{Kinematic information of HH objects (Continued)}\\
\hline
HHO & Proper motion & P.A. & Shock Velocity & Radial Velocity & Reference \\
 & (mas $\mathrm{yr^{-1}}$) & (degree) & (km $\mathrm{s^{-1}}$) & (km $\mathrm{s^{-1}}$) & \\
\hline
\endhead
\hline
\endfoot
\hline
\endlastfoot
        96 & - & - & - & -52 & \citep{wang2004optical} \\ 
        97 & - & - & - & -125 & \citep{wang2004optical} \\ 
        100 & - & - & - & -133 & \citep{wang2004optical} \\ 
        104 A & - & - & - & -46 & \citep{wang2004optical} \\ 
        104 B & - & - & - & -56 & \citep{wang2004optical} \\ 
        119 A1 & 241$\pm$10 & 261$\pm$3 & 62.9$\pm$16 & - & \citep{gaalfalk2007herbig} \\ 
        119 A2 & 207$\pm$11 & 248$\pm$3 & - & - & \citep{gaalfalk2007herbig} \\ 
        119 A3 & 190$\pm$10 & 283$\pm$3 & - & - & \citep{gaalfalk2007herbig} \\ 
        119 B & 278$\pm$11 & 255$\pm$3 & 36.9$\pm$1.5 & - & \citep{gaalfalk2007herbig} \\ 
        119 C & 274$\pm$18 & 100$\pm$4 & 34.8$\pm$6.4 & - & \citep{gaalfalk2007herbig} \\ 
        119 D & 200$\sim$280 & - & - & - & \citep{gaalfalk2007herbig} \\ 
        119 F & 260$\pm$53 & 100$\pm$10 & - & - & \citep{gaalfalk2007herbig} \\ 
        120 A/B/C/D & 26 & 332 & - & - & \citep{kajdic2010herbig} \\ 
        120 E & 45 & 309 & - & - & \citep{kajdic2010herbig} \\ 
        158 C & 197 & 223 & - & - & \citep{mcgroarty2007proper} \\ 
        201 & - & - & - & -284$\sim$-260 & \citep{doi2004internal} \\ 
        113-153 HH201 & 179 & 320 & - & - & \citep{bally2000disks} \\ 
        112-152 HH 201 & 173 & 33 & 321 & -270 & \citep{doi2004internal} \\ 
        115-155 HH201 & 43 & 320 & - & - & \citep{bally2000disks} \\ 
        113-155 HH 201 & 47 & 9 & 298 & -294 & \citep{doi2004internal} \\ 
        116-156 HH201 & 80 & 320 & - & - & \citep{bally2000disks} \\ 
        202 & 22$\sim$118 & 255$\sim$3 & - & -68$\sim$-9 & \citep{o2003high, doi2004internal} \\ 
        117-256 HH 202 & 59 & 41 & 89 & -67 & \citep{doi2004internal} \\ 
        117-256 HH 202 & 59 & 53 & 74 & -44 & \citep{doi2004internal} \\ 
        203 & - & - & - & -69$\sim$-14 & \citep{doi2004internal} \\ 
        221-501 HH 203 & 73 & 45 & 104 & -74 & \citep{doi2004internal} \\ 
        222-506 HH203 & 73$\pm$20 & 140 & - & - & \citep{bally2000disks} \\ 
        222-505 HH203 & 71$\pm$20 & 140 & - & - & \citep{bally2000disks} \\ 
        223-458 HH203 & 61$\pm$20 & 140 & - & - & \citep{bally2000disks} \\ 
        223-500 HH203 & 77$\pm$20 & 140 & - & - & \citep{bally2000disks} \\ 
        223-504 HH203 & 59$\pm$20 & 140 & - & - & \citep{bally2000disks} \\ 
        204 & - & - & - & -68$\sim$-8 & \citep{doi2004internal} \\ 
        226-516 HH 204 & 92 & 63 & 103 & -46 & \citep{doi2004internal} \\ 
        226-517 HH204 & 59$\pm$10 & 140 & - & - & \citep{bally2000disks} \\ 
        228-520 HH204 & 62$\pm$10 & 140 & - & - & \citep{bally2000disks} \\ 
        117-025 HH205 & 346$\pm$10 & 330 & - & - & \citep{bally2000disks} \\ 
        120-035 HH205 & 300$\pm$5 & 330 & - & - & \citep{bally2000disks} \\ 
        121-034 HH205 & 164$\pm$10 & 330 & - & - & \citep{bally2000disks} \\ 
        220 NW & 323 & 287$\pm$10 & - & - & \citep{mcgroarty2007proper} \\ 
        222 & 107$\pm$25 & 329$\pm$10 & - & -6$\pm$10 & \citep{reipurth2013hh} \\ 
        228 E2 & 357 & - & - & - & \citep{wang2009herbig} \\ 
        228 E & 335 & - & - & - & \citep{wang2009herbig} \\ 
        228 W & 264 & - & - & - & \citep{wang2009herbig} \\ 
        230 & - & 70 & 50 & - & \citep{mcgroarty2004classical} \\ 
        269 & 11$\sim$101 & 274$\sim$293 & - & - & \citep{o2003high} \\ 
        079-345 HH269 & 76 & 260 & - & - & \citep{bally2000disks} \\ 
        097-347 HH269 & 60 & 250 & - & - & \citep{bally2000disks} \\ 
        280 A & - & - & - & -45$\pm$10 & \citep{walawender2004deep} \\ 
        317 A & - & - & - & -15$\pm$10 & \citep{walawender2004deep} \\ 
        317 B & - & - & - & $+$15$\pm$10 & \citep{walawender2004deep} \\ 
        317 C & - & - & - & $+$35$\pm$10 & \citep{walawender2004deep} \\ 
        366 & - & - & - & $+$42$\pm$3 & \citep{stecklum2004high} \\ 
        400 W1 & - & - & 30 & 12 & \citep{bally2001kinematics} \\ 
        400 W2 & - & - & 30 & 14 & \citep{bally2001kinematics} \\ 
        400 W3 & - & - & 30 & 9 & \citep{bally2001kinematics} \\ 
        400 W4 & - & - & 30 & 6 & \citep{bally2001kinematics} \\ 
        400 W5 & - & - & 30 & 13 & \citep{bally2001kinematics} \\ 
        400 W6 & - & - & 30 & 15 & \citep{bally2001kinematics} \\ 
        400 W7 & - & - & 30 & 14 & \citep{bally2001kinematics} \\ 
        400 Tip1 & - & - & 30 & 12 & \citep{bally2001kinematics} \\ 
        400 Tip2 & - & - & 30 & 12 & \citep{bally2001kinematics} \\ 
        400 Clump1 & - & - & 30 & 8 & \citep{bally2001kinematics} \\ 
        400 Clump2 & - & - & 30 & 4 & \citep{bally2001kinematics} \\ 
        400 E1 & - & - & 30 & 6 & \citep{bally2001kinematics} \\ 
        400 E2 & - & - & 30 & 4 & \citep{bally2001kinematics} \\ 
        400 E3 & - & - & 30 & 5 & \citep{bally2001kinematics} \\ 
        627 A–D & - & - & 200 & - & \citep{magakian2010wide} \\ 
        450 & - & - & - & -25$\sim$-50 & \citep{bally2001star} \\ 
        492 & - & - & - & $+$95$\pm$10 & \citep{walawender2004deep} \\ 
        493 & - & - & - & $+$65$\pm$10 & \citep{walawender2004deep} \\ 
        494 & - & - & - & -45$\pm$7 & \citep{stecklum2004high} \\ 
        502 NE & 245.7$\pm$10.9 & 32.7 & - & - & \citep{bally2006irradiated} \\ 
        502 N1 & 186.1$\pm$28.5 & 18.5 & - & - & \citep{bally2006irradiated} \\ 
        502 N1 & 197.4$\pm$31.2 & 14.5 & - & - & \citep{bally2006irradiated} \\ 
        502 N1 & 177.8$\pm$13.3 & 61.4 & - & - & \citep{bally2006irradiated} \\ 
        502 N2 & 46.6$\pm$16.7 & 37.3 & - & - & \citep{bally2006irradiated} \\ 
        502 N2 & 32.6$\pm$12.9 & 358.6 & - & - & \citep{bally2006irradiated} \\ 
        502 N2 & 58.3$\pm$7.5 & 39.7 & - & - & \citep{bally2006irradiated} \\ 
        502 N2 & 59.1$\pm$10.4 & 38.5 & - & - & \citep{bally2006irradiated} \\ 
        502 N3 & 36.4$\pm$10.4 & 41.5 & - & - & \citep{bally2006irradiated} \\ 
        502 N3 & 25.1$\pm$15.6 & 41.3 & - & - & \citep{bally2006irradiated} \\ 
        502 N3 & 15.2$\pm$12.3 & 40 & - & - & \citep{bally2006irradiated} \\ 
        502 N3 &  4.8$\pm$28.4 & 44.6 & - & - & \citep{bally2006irradiated} \\ 
        502 SW & 138.0$\pm$9.1 & 208 & - & - & \citep{bally2006irradiated} \\ 
        502 S4 & 147.3$\pm$18.5 & 206.6 & - & - & \citep{bally2006irradiated} \\ 
        502 S5 & 106.9$\pm$34.9 & 206.1 & - & - & \citep{bally2006irradiated} \\ 
        502 S6 & 75.4$\pm$12.2 & 204.6 & - & - & \citep{bally2006irradiated} \\ 
        502 S6 & 112.6$\pm$27.3 & 204.1 & - & - & \citep{bally2006irradiated} \\ 
        502 S7 & 148.3$\pm$38.0 & 206.9 & - & - & \citep{bally2006irradiated} \\ 
        502 S7 & 121.3$\pm$58.5 & 176.6 & - & - & \citep{bally2006irradiated} \\ 
        505 N1 & 144.3$\pm$12.6 & 338.4 & - & - & \citep{bally2006irradiated} \\ 
        505 N1 & 128.6$\pm$11.2 & 340.9 & - & - & \citep{bally2006irradiated} \\ 
        505 N1 & 139.9$\pm$7.9 & 330.4 & - & - & \citep{bally2006irradiated} \\ 
        505 N1 & 150.7$\pm$15.2 & 314.7 & - & - & \citep{bally2006irradiated} \\ 
        505 N1 & 120.0$\pm$16.4 & 323.2 & - & - & \citep{bally2006irradiated} \\ 
        505 N2 & 89.4$\pm$14.2 & 321.6 & - & - & \citep{bally2006irradiated} \\ 
        505 & 109.2$\pm$23.5 & 298.7 & - & - & \citep{bally2006irradiated} \\ 
        505 N3 & 42.8$\pm$30.8 & 278.9 & - & - & \citep{bally2006irradiated} \\ 
        505 jet & 155.9$\pm$25.8 & 187.5 & - & - & \citep{bally2006irradiated} \\
        505 S1 & 124.7$\pm$24.5 & 188.6 & - & - & \citep{bally2006irradiated} \\ 
        505 & 147.4$\pm$19.7 & 200.1 & - & - & \citep{bally2006irradiated} \\ 
        505 & 93.9$\pm$9.3 & 233.1 & - & - & \citep{bally2006irradiated} \\ 
        505 & 49.8$\pm$28.7 & 269.6 & - & - & \citep{bally2006irradiated} \\ 
        109-347 HH507 & 39 & 315 & - & - & \citep{bally2000disks} \\ 
        507 & 3$\sim$43 & 218$\sim$341 & - & - & \citep{o2003high} \\ 
        136-301 HH509 & 42 & 230 & - & - & \citep{bally2000disks} \\ 
       512 & - & - & - & $+$48$\sim$$+$116 & \citep{doi2004internal} \\ 
       178-514 HH 512 & 130 & 113 & 141 & $+$54 & \citep{doi2004internal} \\ 
       157-237 HH513 & 30$\pm$15 & 260 & - & - & \citep{bally2000disks} \\ 
       159-237 HH513 & 64$\pm$5 & 260 & - & - & \citep{bally2000disks} \\ 
       161-236 HH513 & 64$\pm$30 & 260 & - & - & \citep{bally2000disks} \\ 
       162-236 HH513 & 104$\pm$10 & 260 & - & - & \citep{bally2000disks} \\ 
       168-235 HH513 & 51$\pm$10 & 90 & - & - & \citep{bally2000disks} \\ 
       172-234 HH513 & 59$\pm$20 & 80 & - & - & \citep{bally2000disks} \\ 
       170-334 HH514 & 61 & 360 & - & - & \citep{bally2000disks} \\ 
       514 & - & - & - & $+$136 & \citep{doi2004internal} \\ 
       169-333 HH 514 & 37 & 164 & 131 & $+$126 & \citep{doi2004internal} \\ 
       137-508 HH516 & 28 & 100 & - & - & \citep{bally2000disks} \\ 
       516 & - & - & - & -96$\sim$-91 & \citep{doi2004internal} \\ 
       136-508 HH 516 & 26 & 14 & 106 & -103 & \citep{doi2004internal} \\ 
       164-342 HH518 & 21$\pm$10 & 30 & - & - & \citep{bally2000disks} \\ 
       518 & 2$\sim$28 & 357$\sim$85 & - & $+$52$\sim$$+$98 & \citep{o2003high, doi2004internal} \\ 
       161-344 HH 518 & 19 & 166 & 76 & $+$74 & \citep{doi2004internal} \\ 
       165-333 HH 518 & 21 & 156 & 51 & $+$46 & \citep{doi2004internal} \\ 
       171-326 HH 518 & 42 & 154 & 96 & $+$86 & \citep{doi2004internal} \\ 
       523 & - & - & - & -63$\sim$-43 & \citep{doi2004internal} \\ 
       183-330 HH 523 & 71 & 44 & 102 & -73 & \citep{doi2004internal} \\ 
       179-326 HH523 & 75 & 100 & - & - & \citep{bally2000disks} \\ 
       183-327 HH523 & 96 & 100 & - & - & \citep{bally2000disks} \\ 
       184-330 HH523 & 66 & 90 & - & - & \citep{bally2000disks} \\ 
       528 & 4$\sim$53 & 121$\sim$222 & - & - & \citep{o2003high} \\ 
       162-414 HH528 & 41 & 190 & - & - & \citep{bally2000disks} \\ 
       165-436 HH528 & 31 & 150 & - & - & \citep{bally2000disks} \\ 
       166-441 HH528 & 39 & 110 & - & - & \citep{bally2000disks} \\ 
       167-434 HH528 & 21 & 130 & - & - & \citep{bally2000disks} \\ 
       167-433 HH528 & 21 & 130 & - & - & \citep{bally2000disks} \\ 
       168-439 HH528 & 30 & 130 & - & - & \citep{bally2000disks} \\ 
       176-455 HH528 & 39 & 160 & - & - & \citep{bally2000disks} \\ 
       179-515 HH528 & 65 & 75 & - & - & \citep{bally2000disks} \\ 
       181-510 HH528 & 26 & 180 & - & - & \citep{bally2000disks} \\ 
       182-506 HH528 & 30 & 180 & - & - & \citep{bally2000disks} \\ 
       182-510 HH528 & 21 & 165 & - & - & \citep{bally2000disks} \\ 
       185-511 HH528 & 26 & 170 & - & - & \citep{bally2000disks} \\ 
       193-457 HH528 & 30 & 130 & - & - & \citep{bally2000disks} \\ 
       529 & 21$\sim$114 & 77$\sim$144 & - & -66$\sim$-28 & \citep{o2003high, doi2004internal} \\ 
       151-353 HH 529 & 75$\pm$20 & 100 & - & - & \citep{bally2000disks} \\ 
       149-352 HH 529 & 85 & 49 & 113 & -75 & \citep{doi2004internal} \\ 
       152-354 HH 529 & 88$\pm$30 & 100 & - & - & \citep{bally2000disks} \\ 
       156-353 HH 529 & 131 & 62 & 149 & -71 & \citep{doi2004internal} \\ 
       156-355 HH 529 & 139$\pm$15 & 100 & - & - & \citep{bally2000disks} \\ 
       159-353 HH 529 & 80 & 56 & 97 & -55 & \citep{doi2004internal} \\ 
       156-356 HH 529 & <30 & - & - & - & \citep{bally2000disks} \\ 
       163-403 HH 529 & 67 & 60 & 77 & -38 & \citep{doi2004internal} \\ 
       161-354 HH 529 & 72$\pm$30 & 100 & - & - & \citep{bally2000disks} \\ 
       167-359 HH 529 & 54 & 45 & 76 & -54 & \citep{doi2004internal} \\ 
       158-356 HH 529 & 50$\pm$25 & 250 & - & - & \citep{bally2000disks} \\ 
       170-358 HH 529 & 82 & 100 & - & - & \citep{bally2000disks} \\ 
       172-352 HH 529 & 30 & 60 & - & - & \citep{bally2000disks} \\ 
       530 & 25$\sim$70 & 214$\sim$273 & - & - & \citep{o2003high} \\ 
       105-417 HH530 & 36 & 250 & - & - & \citep{bally2000disks} \\ 
       108-430 HH530 & 33 & 250 & - & - & \citep{bally2000disks} \\ 
       109-416 HH530 & 55 & 260 & - & - & \citep{bally2000disks} \\ 
       540 A & - & - & - & -33 & \citep{bally2001kinematics} \\ 
       540 B & - & - & - & -50 & \citep{bally2001kinematics} \\ 
       540 C & - & - & - & -40 & \citep{bally2001kinematics} \\ 
       555 & - & - & - & -10$\sim$-85 & \citep{bally2003irradiated} \\ 
       558 & - & - & - & 25 & \citep{bally2001kinematics} \\ 
       559 knot1 & - & - & - & -29 & \citep{bally2001kinematics} \\ 
       559 knot2 & - & - & - & -51 & \citep{bally2001kinematics} \\ 
       559 knot3 & - & - & - & -77 & \citep{bally2001kinematics} \\ 
       560 north & - & - & - & 12 & \citep{bally2001kinematics} \\ 
       560 south & - & - & - & -25 & \citep{bally2001kinematics} \\ 
       561 jet & - & - & - & 50 & \citep{bally2001kinematics} \\ 
       586 & - & - & 200 & - & \citep{ogura2002halpha} \\ 
       624 & 19$\sim$35 & 298$\sim$323 & - & - & \citep{o2003high} \\ 
       625 & 25$\sim$31 & 288$\sim$315 & - & - & \citep{o2003high} \\ 
       626 & 15$\sim$43 & 216$\sim$14 & - & -34$\sim$-30 & \citep{o2003high, doi2004internal} \\ 
       141-424 HH 626 & 42 & 46 & 58 & -40 & \citep{doi2004internal} \\ 
       143-427 HH 626 & 36 & 39 & 57 & -44 & \citep{doi2004internal} \\ 
       628 A–C & - & - & 200 & - & \citep{magakian2010wide} \\ 
       628 & 40$\sim$50 & - & - & - & \citep{movsessian2003optical} \\ 
       629 A, B, E–G, X(?) & - & - & 200 & - & \citep{magakian2010wide} \\ 
       630 & - & - & 200 & - & \citep{magakian2010wide} \\ 
       632 & - & - & 200 & - & \citep{magakian2010wide} \\ 
       633 & - & 350 & 200 & - & \citep{magakian2010wide, movsessian2003optical} \\ 
       635 A-H & - & - & 200 & - & \citep{magakian2010wide} \\ 
       666 & 34$\sim$184 & - & - & - & \citep{reiter2017proper} \\ 
       666 D & - & 293.5 & - & -11 & \citep{smith2004hh} \\ 
       666 A & - & 293.5 & - & -37 & \citep{smith2004hh} \\ 
       666 E & - & 293.5 & - & -130 & \citep{smith2004hh} \\ 
       666 M & - & 293.5 & 200 & -190 & \citep{smith2004hh} \\ 
       666 O & - & 293.5 & - & $+$210 & \citep{smith2004hh} \\ 
       666 N & - & 293.5 & - & $+$93 & \citep{smith2004hh} \\ 
       666 I & - & 293.5 & - & $+$67 & \citep{smith2004hh} \\ 
       666 C & - & 293.5 & - & $+$67 & \citep{smith2004hh} \\ 
       668 N3 & 84.5$\pm$21.8 & 354.1 & - & - & \citep{smith2005new} \\ 
       668 N2 & 275.1$\pm$8.4 & 340.3 & - & - & \citep{smith2005new} \\ 
       668 N1 & 275.1$\pm$8.4 & 340.5 & - & - & \citep{smith2005new} \\ 
       668 jet & 303.5$\pm$8.0 & 170 & - & - & \citep{smith2005new} \\ 
       668 S1 & 245.6$\pm$6.0 & 160.2 & - & - & \citep{smith2005new} \\ 
       668 S2 & 108.8$\pm$16.0 & 145.3 & - & - & \citep{smith2005new} \\ 
       668 S3 & 178.2$\pm$16.0 & 159.7 & - & - & \citep{smith2005new} \\ 
       668 B & 111.8$\pm$14.0 & 160.4 & - & - & \citep{smith2005new} \\ 
       668 A & 143.3$\pm$8.4 & 154.3 & - & - & \citep{smith2005new} \\ 
       670 & - & - & 200 & - & \citep{magakian2010wide} \\ 
       702 A & 203, 129 & 225$\pm$(3$\sim$8), 194$\pm$(5$\sim$12) & - & - & \citep{mcgroarty2007proper} \\ 
       702 B & 97 & 285$\pm$(3$\sim$8) & - & - & \citep{mcgroarty2007proper} \\ 
       702 C & 204, 149 & 224$\pm$(6$\sim$17), 128$\pm$(4$\sim$11) & - & - & \citep{mcgroarty2007proper} \\ 
       702 D & 219, 190 & 205$\pm$(3$\sim$8), 269$\pm$(3$\sim$9) & - & - & \citep{mcgroarty2007proper} \\ 
       702 E & 125, 186 & 191$\pm$(3$\sim$7), 117$\pm$(3$\sim$9) & - & - & \citep{mcgroarty2007proper} \\ 
       703 & 170$\pm$30 & 195$\pm$10 & - & - & \citep{bally2012deep} \\ 
       705 & - & 36 & 50 & - & \citep{mcgroarty2004classical} \\ 
       705 A1 & 187, 330 & 180$\pm$(3$\sim$9), 186$\pm$(2$\sim$5) & - & - & \citep{mcgroarty2007proper} \\ 
       705 A2 & 147, 231 & 218$\pm$(4$\sim$11), 186$\pm$(3$\sim$7) & - & - & \citep{mcgroarty2007proper} \\ 
       705 A3 & 244, 194 & 241$\pm$(3$\sim$7), 131$\pm$(3$\sim$8) & - & - & \citep{mcgroarty2007proper} \\ 
       705 A4 & 99, 211 & 232$\pm$(6$\sim$16), 227$\pm$(3$\sim$8) & - & - & \citep{mcgroarty2007proper} \\ 
       777 & 105 & - & - & - & \citep{reipurth2003blowout} \\ 
       826 C & 261, 111 & 321$\pm$12, 26$\pm$29 & - & - & \citep{mcgroarty2007proper} \\ 
       826 A & 132 & 159$\pm$24 & - & - & \citep{mcgroarty2007proper} \\ 
       826 B & 207, 282 & 156$\pm$15, 166$\pm$11 & - & - & \citep{mcgroarty2007proper} \\ 
       827 A & 347 & 180$\pm$(2$\sim$5) & - & - & \citep{mcgroarty2007proper} \\ 
       827 B & 199, 151 & 198$\pm$(3$\sim$8), 188$\pm$(4$\sim$11) & - & - & \citep{mcgroarty2007proper} \\ 
       828 M & 118 & 144$\pm$27 & - & - & \citep{mcgroarty2007proper} \\ 
       828 W & 173 & 208$\pm$18 & - & - & \citep{mcgroarty2007proper} \\ 
       829 A & 116 & 108$\pm$(5$\sim$14) & - & - & \citep{mcgroarty2007proper} \\ 
       829 B & 113 & 163$\pm$(5$\sim$14) & - & - & \citep{mcgroarty2007proper} \\ 
       830 & - & 50 & 150 & - & \citep{mcgroarty2004classical} \\ 
       830 C West & 298 & 314$\pm$28 & - & - & \citep{mcgroarty2007proper} \\ 
       830 C East & 161, 348 & 226$\pm$112, 314$\pm$7 & - & - & \citep{mcgroarty2007proper} \\ 
       830 B & 142, 327 & 313$\pm$22, 233$\pm$7 & - & - & \citep{mcgroarty2007proper} \\ 
       831 & - & 74 & 50 & - & \citep{mcgroarty2004classical} \\ 
       831 A1 & 184 & 331$\pm$(3$\sim$9) & - & - & \citep{mcgroarty2007proper} \\ 
       831 A2 & 166 & 300$\pm$(4$\sim$10) & - & - & \citep{mcgroarty2007proper} \\ 
       831 A3 & 196 & 288$\pm$(3$\sim$8) & - & - & \citep{mcgroarty2007proper} \\ 
       831 A4 & 141 & 32$\pm$(4$\sim$11) & - & - & \citep{mcgroarty2007proper} \\ 
       831 A5 & 112, 221 & 284$\pm$(5$\sim$14), 182$\pm$(3$\sim$7) & - & - & \citep{mcgroarty2007proper} \\ 
       831 B1 & 71 & 158$\pm$(8$\sim$22) & - & - & \citep{mcgroarty2007proper} \\ 
       831 B2 & 79 & 134$\pm$(8$\sim$20) & - & - & \citep{mcgroarty2007proper} \\ 
       832 & - & 78 & 50 & - & \citep{mcgroarty2004classical} \\ 
       833 & - & 25 & 50 & - & \citep{mcgroarty2004classical} \\ 
       874 jet & 165.6$\pm$17.7 & 140.9 & - & - & \citep{bally2006irradiated} \\ 
       874 Small bow & 200.7$\pm$23.3 & 160.9 & - & - & \citep{bally2006irradiated} \\ 
       874 Knot & 144.3$\pm$27.0 & 147.6 & - & - & \citep{bally2006irradiated} \\ 
       874 Large bow & 113.3$\pm$44.2 & 171.1 & - & - & \citep{bally2006irradiated} \\ 
       874 Large bow & 97.0$\pm$37.6 & 125.8 & - & - & \citep{bally2006irradiated} \\ 
       876 & 47.3$\pm$14.6 & 279.9 & - & - & \citep{bally2006irradiated} \\ 
       876 & 58.6$\pm$9.2 & 269.1 & - & - & \citep{bally2006irradiated} \\ 
       876 & 37.3$\pm$16.1 & 233.1 & - & - & \citep{bally2006irradiated} \\ 
       877 & 261.4$\pm$14.3 & 44.6 & - & - & \citep{bally2006irradiated} \\ 
       896 & - & - & 300 & - & \citep{barba2007geysers} \\ 
       897 & - & 150 & 300 & - & \citep{barba2007geysers} \\ 
       899 & - & - & - & -160$\sim$-185 & \citep{walawender2013optical} \\ 
       900 & 37$\sim$104 & - & 200 & - & \citep{reiter2017proper, smith2010hst} \\ 
       900 A & 60 & - & 60 & - & \citep{reiterDisentanglingOutflowProtostars2015} \\ 
       900 microjet & - & - & 200 & - & \citep{smith2010hst} \\ 
       900 B & 37 & - & 37 & - & \citep{reiterDisentanglingOutflowProtostars2015} \\ 
       900 C & 68 & - & 68 & - & \citep{reiterDisentanglingOutflowProtostars2015} \\ 
       900 D & 97 & - & 97 & - & \citep{reiterDisentanglingOutflowProtostars2015} \\ 
       900 E & 97 & - & 97 & - & \citep{reiterDisentanglingOutflowProtostars2015} \\ 
       900 F & 89 & - & 90 & -16 & \citep{reiterDisentanglingOutflowProtostars2015} \\ 
       900 G & 104 & - & 105 & -7 & \citep{reiterDisentanglingOutflowProtostars2015} \\ 
       900 H & 51 & - & 51 & - & \citep{reiterDisentanglingOutflowProtostars2015} \\ 
       900 I & 61 & - & 61 & - & \citep{reiterDisentanglingOutflowProtostars2015} \\ 
       901 & 51$\sim$94 & - & - & - & \citep{reiter2017proper} \\ 
       901 E & - & - & 200 & - & \citep{smith2010hst} \\ 
       901 W & - & - & 200 & - & \citep{smith2010hst} \\ 
       902 W & - & - & 200 & - & \citep{smith2010hst} \\ 
       902 & 35$\sim$161 & - & - & - & \citep{reiter2017proper} \\ 
       903 jet & - & - & 200 & - & \citep{smith2010hst} \\ 
       903 & 40$\sim$140 & - & - & - & \citep{reiter2017proper} \\ 
       948 B & 31 & 308 & - & - & \citep{kajdic2010herbig} \\ 
       950 A & 82 & 253 & - & - & \citep{kajdic2010herbig} \\ 
       950 C & 89 & 240 & - & - & \citep{kajdic2010herbig} \\ 
       950 D & 38 & 250 & - & - & \citep{kajdic2010herbig} \\ 
       950 E & 107 & 244 & - & - & \citep{kajdic2010herbig} \\ 
       966 A–D & - & - & 200 & - & \citep{magakian2010wide} \\ 
       967 & - & - & 200 & - & \citep{magakian2010wide} \\ 
       968 A–G & - & - & 200 & - & \citep{magakian2010wide} \\ 
       973 & - & - & 200 & - & \citep{magakian2010wide} \\ 
       974 A–B & - & - & 200 & - & \citep{magakian2010wide} \\ 
       975 A–B & - & - & 200 & - & \citep{magakian2010wide} \\ 
       991 & 150 & - & - & - & \citep{wang2009herbig} \\ 
       1002 A & 25 & 17 & - & - & \citep{reiter2022deep} \\ 
       1002 B & 50 & 107 & - & - & \citep{reiter2022deep} \\ 
       1002 C & 57 & 286 & - & - & \citep{reiter2022deep} \\ 
       1003 A & - & - & 200 & - & \citep{smith2010hst} \\ 
        1003 A & 115 & 182 & - & - & \citep{reiter2022deep} \\ 
       1003 B & 115 & 165 & - & - & \citep{reiter2022deep} \\ 
       1003 C & 50 & 160 & - & - & \citep{reiter2022deep} \\ 
       1004 NE & - & - & 200 & - & \citep{smith2010hst} \\ 
       1004 & 33$\sim$183 & - & - & - & \citep{reiter2017proper} \\ 
       1005 & 1$\sim$162 & - & 200 & - & \citep{smith2010hst,reiter2017proper} \\ 
       1006 & 42$\sim$156 & - & - & - & \citep{reiter2017proper} \\ 
       1006 N & - & - & 200 & - & \citep{smith2010hst} \\ 
       1006 S & - & - & 200 & - & \citep{smith2010hst} \\ 
       1007 & 40$\sim$77 & - & 200 & - & \citep{smith2010hst, reiter2017proper} \\ 
       1008 & 45$\sim$67 & - & - & - & \citep{reiter2017proper} \\ 
       1009 & 46$\sim$122 & - & - & - & \citep{reiter2017proper} \\ 
       1010 SW & - & - & 200 & - & \citep{smith2010hst} \\ 
       1010 & 35$\sim$173 & - & - & - & \citep{reiter2017proper} \\ 
       1011 & 0.5$\sim$9 & - & 200 & - & \citep{smith2010hst, reiter2017proper} \\ 
       1012 microjet & - & - & 200 & - & \citep{smith2010hst} \\ 
       1012 & 65$\sim$181 & - & - & - & \citep{reiter2017proper} \\ 
       1013 & 73$\sim$155 & - & 200 & - & \citep{smith2010hst,reiter2017proper} \\ 
       1014 & 30$\sim$93 & - & 200 & - & \citep{smith2010hst, reiter2017proper} \\ 
       1015 & 74$\sim$107 & - & 200 & - & \citep{smith2010hst, reiter2017proper} \\ 
       1016 & 74$\sim$107 & - & 200 & - & \citep{smith2010hst, reiter2017proper} \\ 
       1017 jet & - & - & 200 & - & \citep{smith2010hst} \\ 
       1017 & 4$\sim$ 98 & - & - & - & \citep{reiter2017proper} \\ 
       1018 microjet & - & - & 200 & - & \citep{smith2010hst} \\ 
       1018 & 38$\sim$94 & - & - & - & \citep{reiter2017proper} \\ 
       1019 & 50$\sim$100 & - & - & - & \citep{reiter2017dusty} \\ 
       1019 & 43$\sim$115 & - & - & - & \citep{reiter2017proper} \\ 
       1066 & 41$\sim$208 & - & - & - & \citep{reiter2017proper} \\ 
       1156 & 41$\sim$85 & - & - & - & \citep{reiter2017proper} \\ 
       1158 & - & - & - & -30.9 & \citep{riaz2015hh} \\ 
       1159 & 31$\sim$65 & - & - & - & \citep{reiter2017proper} \\ 
       c-5(1160) & 200 & 190 & - & - & \citep{reiter2022deep} \\ 
       1160 & 65$\sim$75 & - & - & - & \citep{reiter2017proper} \\ 
       1161 & 32$\sim$123 & - & - & - & \citep{reiter2017proper} \\ 
       1162 & 28$\sim$65 & - & - & - & \citep{reiter2017proper} \\ 
       1163 & 62 & - & - & - & \citep{reiter2017proper} \\ 
       1164 & 48$\sim$54 & - & - & - & \citep{reiter2017proper} \\ 
       1165 & 80$\pm$10 & 318$\pm$5 & - & - & \citep{riaz2017first} \\ 
       1166 & 61 & - & - & - & \citep{reiter2017proper} \\ 
       1167 & 65 & - & - & - & \citep{reiter2017proper} \\ 
       1168 & 41 & - & - & - & \citep{reiter2017proper} \\ 
       1169 & 69 & - & - & - & \citep{reiter2017proper} \\ 
       1170 & 84 & - & - & - & \citep{reiter2017proper} \\ 
       1171 & 34$\sim$174 & - & - & - & \citep{reiter2017proper} \\ 
       1172 & 235 & - & - & - & \citep{reiter2017proper} \\ 
       1173 & 58$\sim$59 & - & - & - & \citep{reiter2017proper} \\ 
       1181 A & - & - & - & -160$\pm$5 & \citep{dodin2024jet} \\ 
       1181 B & - & - & - & $+$138$\pm$7 & \citep{dodin2024jet} \\ 
       1218 & 70 & 295 & - & - & \citep{reiter2022deep} \\ 
       1218 counterjet & 100 & 112 & - & - & \citep{reiter2022deep} \\ 
       1219 & 85, 100 & 139 & - & - & \citep{reiter2022deep} \\ 
       1220 & 130 & 185 & - & - & \citep{reiter2022deep} \\ 
       1221 east & 100 & 355 & - & - & \citep{reiter2022deep} \\ 
       1222 & 100 & 119 & - & - & \citep{reiter2022deep} \\ 
       1223 ridges & 35 & 110 & - & - & \citep{reiter2022deep} \\ 
       1223 left/slow & 50 & 105 & - & - & \citep{reiter2022deep} \\ 
       1224 & 140 & 152 & - & - & \citep{reiter2022deep} \\ 
       1225 & 100 & 66 & - & - & \citep{reiter2022deep} \\ 
\end{longtable}
\end{landscape}
\section{The full exciting source catalogue of HH objects}
\label{sec:tab_host}
\begin{longtable}{lll}
\caption{Exciting sources of HH objects}\label{tab:hosts}\\
\hline
HH Objects & Exciting source & Source type \\
\hline
\endfirsthead
\caption[]{Exciting sources of HH objects (Continued)}\\
\hline
HH Objects & Exciting source & Source type \\
\hline
\endhead
\hline
\endfoot
\hline
\endlastfoot
434   & IRAS 04325+2402                                            & -    \\
435   & IRAS 04325+2402                                            & -    \\
436   & IRAS 04325+2402                                            & -    \\
439   & IRAS 06045-0554                                            & A0 D  \\
450   & IRAS 22129+7000                                            & -  \\
462   & IRAS 03507+3801                                            & F5-K0 E \\
463   & IRAS 04073+3800                                            &  -   \\
464   & IRAS 04073+3800                                            &  -   \\
465   & IRAS 04073+3800                                            &  -   \\
466   & IRAS 04305+2414                                            &   \\
467   & IRAS 04305+2414                                            & K5Ve C \\
468   & IRAS 04305+2414                                            & K5Ve C  \\
469   & IRAS 05369-0728                                            & K0 D \\
470   & IRAS 05369-0728                                            & K0 D  \\
471   & IRAS 05451+0037                                            & -  \\
472   & IRAS 05451+0037                                            &  -   \\
473   & IRAS 05451+0037                                            &  -   \\
474   & IRAS 05451+0037                                            &  -   \\
487   & IRAS 02224+7227                                            &  -   \\
488   & IRAS 02238+7222                                            &  -   \\
489   & IRAS 02249+7230                                            &  -  \\
490   & IRAS 05487+0255                                            & -    \\
491   & IRAS 05487+0255                                            & -    \\
492   & IRAS 03235+3004                                            &  -   \\
493   & IRAS 03235+3004                                            & -  \\
494   & CB 26 YSO 1                                                &  -   \\
528   & Source 145-356                                             &  -   \\
529   & Source 144-351                                             &  -  \\
530   & Source 136-355                                             & -  \\
546   & IRAS 4B                                                    & -  \\
549   & IRAS 16226-2319                                            & -    \\
553   & IRAS 16293-2422                                            & -    \\
554   & IRAS 16293-2422                                            & -    \\
563   & IRAS 20489+4406                                            &  -   \\
564   & IRAS 20489+4406                                            &  -   \\
565   & IRAS 20489+4406                                            &  -   \\
567   & IRAS 20489+4410                                            &  -  \\
571   & IRAS 06382+1017                                            &  -  \\
572   & IRAS 06386+1023                                            &  -  \\
576   & IRAS 06381+1039                                            &  -   \\
577   & IRAS 06381+1039                                            &  -  \\
579   & IRS 1                                                      &  -   \\
580   & KH$\alpha$ 112                                             &  -  \\
586   & IRAS 02252+6120                                            &  - \\
587   & IRAS 20489+4410                                            &  -  \\
588   & IRAS 21388+5622                                            &  -    \\
589   & IRAS 21391+5802                                            & -   \\
594   & IRAS 21391+5802                                            & -    \\
595   & IRAS 21391+5802                                            & -    \\
596   & IRAS 22266+6358                                            &  -  \\
600   & Par-Lup3-4                                                 & F7 D  \\
619   & IRAS 22178+6317                                            & -    \\
620   & IRAS 22178+6317                                            & -    \\
621   & IRAS 22178+6317                                            & -    \\
622   & IRAS 22178+6317                                            & -    \\
624   & Source 136-360                                             & -    \\
625   & Source 136-360                                             & -    \\
626   & Source 136-360                                             & -    \\
627   & CN 6                                                       & [WC] D    \\
628   & CN 3                                                       & [WC8] C   \\
630   & CN 2                                                       & -    \\
632   & CN 1                                                       & -    \\
646   & LkH$\alpha$ 185                                            & K6e D  \\
653   & MKH$\alpha$ 10                                             & -   \\
658   & LkH$\alpha$ 186                                            & M0.6 D  \\
659   & LkH$\alpha$ 187                                            & M1.2 D \\
660   & LkH$\alpha$ 187                                            & M1.2 D  \\
661   & MKH$\alpha$ 24                                             &  -   \\
667   & d216-0939                                                  & -    \\
668   & d253-1536                                                  & -    \\
673   & WL 18, [GY92] 129                                          & M3.5 D  \\
674   & WLY 2-44, YLW 16A, [GY92] 269                              &  K8 D \\
675   & WLY 2-51, YLW 45, [GY92] 315                               &  -   \\
676   & V2129 Oph, YLW 49, WLY 2-52, WSB 54, [GY92] 319 [GY92] 284 & K5e C  \\
677   & V2059 Oph, DoAr 37, WSB 57, YLW 56, [GY92] 400             & M1.5e D \\
696   & IRAS 04325+2402                                            & -   \\
697   & IRAS 04325+2402                                            &  -  \\
698   & IRAS 04325+2402                                            &  -   \\
699   & IRAS 04325+2402                                            &  -   \\
702   & IRAS 04234+2547, IRAS 04237+2559, IRAS 04234+2547         & -    \\
704   & IRAS 04356+2516                                           & -    \\
705   & LDN 1527 1, LDN 1627 2                                    & -    \\
706   & ITG 9, ITG 9B, ITG 9C, IRAS 04356+2516, IRAS 04357+2528   & M0III D   \\
708   & ROXs 20A, ROXs 20B                                        &  M4.5 D, M2 D \\
712   & SR 21, Elias 34                                           &  -   \\
727   & ASR 41                                                     &  -   \\
729   & S CrA                                                      & G0Ve+K0Ve C  \\
730   & IRS 6                                                      & F0 D  \\
733   & T CrA                                                      &  -   \\
801   & LkH$\alpha$ 198                                            & B9e D  \\
802   & LkH$\alpha$ 198                                            & B9e D  \\
803   & 1548C27 IRS 1                                              &  -  \\
804   & IRAS 19395+2313                                            &  -  \\
805   & IRAS 19395+2313                                            &  -  \\
808   & LkH$\alpha$ 233                                            & A7Ve C  \\
809   & LkH$\alpha$ 233                                            & A7Ve C  \\
810   & LkH$\alpha$ 233                                            & A7Ve C  \\
811   & LkH$\alpha$ 233                                            & A7Ve C  \\
812   & LkH$\alpha$ 233                                            & A7Ve C \\
813   & LkH$\alpha$ 233                                            & A7Ve C  \\
814   & LkH$\alpha$ 233                                            & A7Ve C  \\
815   & LkH$\alpha$ 234                                            & B5Ve C \\
816   & LkH$\alpha$ 234                                            & B5Ve C \\
817   & LkH$\alpha$ 234                                            & B5Ve C\\
818   & LkH$\alpha$ 234                                            & B5Ve C \\
819   & LkH$\alpha$ 234                                            & B5Ve C \\
820   & LkH$\alpha$ 234                                            & B5Ve C \\
821   & LkH$\alpha$ 234                                            & B5Ve C\\
822   & LkH$\alpha$ 234                                            & B5Ve C \\
824   & IRAS 21416+6556                                            &  -  \\
825   & IRAS 21416+6556                                            &  -  \\
826   & CWTau                                                      & -   \\
827   & CWTau                                                      &  -   \\
829   & CWTau                                                      &  -   \\
830   & DGTau                                                      &  -   \\
831   & DOTau                                                      &  -   \\
832   & DOTau                                                      &  -   \\
833   & HVTau C                                                    &  -   \\
835   & RW Aur                                                     & K1/5e+K5e C \\
836   & DG Tau B                                                   & -  \\
837   & DG Tau B                                                   &  -   \\
860   & IRS 6                                                      &  -   \\
861   & VV CrA                                                     & K7 D  \\
862   & VV CrA                                                     & K7 D  \\
865   & IRAS 21445+5712                                            & -   \\
866   & IRAS 06046-0603                                            &  -   \\
892   & IRAS 04376+5413                                            &  -   \\
896   & IRAS 18014–2428                                            & -  \\
897   & IRAS 18014–2428                                            & -  \\
899   & LkH$\alpha$ 324SE                                          &  -   \\
900   & Tr16                                                       &  -  \\
901   & Tr14                                                       &  -  \\
902   & Tr14                                                       &  -  \\
903   & Pos 23                                                     &  -  \\
906   & CED 110-IRS 4, CHXR15                                     &  -   \\
908   & ESO H$\alpha$ 560                                          &  -   \\
911   & YSO [CCE98] 69, YSO [CCE98] 33, YSO [CCE98] 34            &  -   \\
912   & [CCE98] 44                                                 &  -   \\
913   & undetected                                                 &  -   \\
914   & T Tauri star Sz 32                                         &  -   \\
915   & ISO Cha I 192                                              &  -  \\
916   & [CCE98] 44                                                 & -   \\
918   & [CCE98] 54                                                 &  -   \\
919   & ESO H$\alpha$ 569                                          &  -   \\
920   & HM 16                                                      &  -  \\
921   & CU Cha                                                     & A0Vep C \\
922   & Cha-MMS1                                                   & -   \\
923   & VW Cha                                                     & Kv7+M0 C  \\
924   & Cha-MMS1                                                   &  -   \\
925   & Cha-MMS1                                                   & -   \\
926   & VW Cha                                                     & K7+M0 C  \\
927   & DI Cha                                                     & G2Ve C  \\
928   & CW Cha                                                     &  -   \\
929   & Ced 110 IRS 4                                              &  -   \\
931   & WW Cha /Cha-MMS2                                           &  -   \\
932   & WZ Cha                                                     & M4Ve C \\
933   & HM 23/Cha-MMS2                                             &  -   \\
934   & HM 23/Cha-MMS2                                             &  -   \\
936   & CHXR 15                                                   & M5+M5 C  \\
939   & classical T Tauri star Sz 50                               &  -   \\
957   & MKH$\alpha$ 27                                             &  -   \\
960   & YSO \#22                                                   & -   \\
962   & a highly embedded IR source                                & -   \\
963   & a nake star(05:54:40, +01:52:42)                           & K3e D \\
965   & 965 source                                                 &  -   \\
966   & IRAS 20568+5217                                            &  -   \\
967   & IRAS 20568+5217                                            &  -   \\
968   & CN 9 (IRAS 20573+5221)                                     &  -   \\
973   & CN 5                                                       &  -   \\
974   & CN 4                                                       &  -   \\
975   & IRAS 20583+5228                                            &  -   \\
979   & IRS 7                                                      &  -   \\
981   & YSO 12                                                     &  -   \\
982   & YSO 12                                                     &  -   \\
983   & Sz 69 (YSO 16)                                             &  M1 D  \\
984   & Sz 69 (YSO 16)                                             &  M1 D  \\
985   & YSOs 14 and 17                                            &  -   \\
986   & YSO 13                                                     &  -   \\
987   & YSO 17                                                     &  -   \\
988   & T Tauri star HM Lup                                        & M3e D  \\
990   & Par-Lup3-4                                                 &  -   \\
991   & Par-Lup3-4                                                 &  -   \\
993   & IRAS 05358+3543                                            &  -  \\
994   & IRAS 05358+3543                                            &  -  \\
998   & AC Ori                                                     & F5e D  \\
1000  & IRAS 17130-2053                                            & -    \\
1002  & NGC 3324                                                   &  -  \\
1003  & NGC 3324                                                   &  -  \\
1004  & Pos 21                                                     &  -  \\
1005  & Pos 21                                                     & -  \\
1006  & Pos 22                                                     &  -  \\
1007  & Pos 25                                                     &  -  \\
1008  & Pos 27                                                     &  -  \\
1009  & Pos 27                                                     &  -  \\
1010  & Pos 30                                                     &  -  \\
1011  & Tr15                                                       & -  \\
1012  & Tr14                                                       & -  \\
1013  & Tr14                                                       &  -  \\
1014  & Tr16                                                       &  -  \\
1015  & Pos 25                                                     &  -  \\
1016  & Pos 24                                                     &  -  \\
1017  & Tr14                                                       &  -  \\
1036  & J20314543+4018526                                          &  -  \\
1040  & IRAS 20305+4010                                            &  -  \\
1042  & 08576nr292                                                 &  -    \\
1043  & 08576nr480                                                 &  -  \\
1090  & ESO-H$\alpha$274                                           &  -  \\
1091  & ESO-H$\alpha$2431                                          &  -   \\
1092  & IRAS08569-4230                                             &  -  \\
1093  & IRAS08569-4230                                             & -  \\
1094  & IRAS08569-4230                                             & -  \\
1095  & BLASTJ090016-443850                                        &  -   \\
1096  & MSXG265.9151+01.0702                                       &  -   \\
1097  & MSXG265.4865+01.3474                                       & -  \\
1098  & IRS35                                                      &  -  \\
1101  & MSXG266.3267+00.9389                                       & -   \\
1107  & HH74IRS                                                    & -   \\
1158  & Mayrit 1082188 (M1082188)                                  & M0.0 D    \\
1165  & Mayrit 1701117 (M1701117)                                  &  -   \\
1181  & BP Tau                                                     & K5/7Ve C    \\
1183  & IRS2, IRS4                                                 &  -   \\
1184  & IRS2, IRS4                                                 &  -   \\
1185  & IRS2, IRS4                                                 &  -   \\
1186  & IRS2, IRS4                                                 &  -   \\
1187  & AS 353                                                     & K5 D \\
1188  & AS 353                                                     & K5 D  \\
1189  & AS 353                                                     & K5 D  \\
1190  & AS 353                                                     & K5 D \\
1191  & AS 353                                                     & K5 D  \\
1192  & IRDC                                                       &  -   \\
1193  & IRDC                                                       & -   \\
1194  & IRDC                                                       &  -   \\
1196  & IRAS 06297+1021 (W) (2MASS J06322611+1019184)              &  -   \\
1197  & 2MASS J06315870+1027474 (IRAS 06292+1029)                  &  -   \\
1201  & LkH$\alpha$ 342 (HBC 204)                                  & M0 D  \\
1202  & VY Mon                                                     & A5:Vep D  \\
1203  & IRAS 06277+1016 (2MASS J06302857+1014236)                  & -   \\
1205  & J085309.91-421232.6 (IRS 26-35)                            &  -   \\
1211  & 2MASS J15003604-6313151                                    & -   \\
1212  & 2MASS J15004103-6306381                                    &  -   \\
1216  & IRAS 06212-1049                                            &  -  \\
1217  & IRAS 06216-1044                                            &  -   \\
1218  & J103654.2-583626                                           & -  \\
1219  & SPICY 7434                                                 &  -  \\
1221  & J103653.8-583748                                           &  -  \\
1226  & IRAS 23591+4748                                            & G7-M4 E  \\
1227  & IRAS 01166+6635                                            &  -   \\
1228  & WISEA J043041.15+352941.4                                  &  -   \\
1229  & 2MASS 06590141-1159424                                     &  -  \\
1230  & IRAS 19219+2300                                            &  -   \\
1231  & IRAS 20472+4338                                            &  -   \\
\end{longtable}

\bibliography{ref}
\bibliographystyle{aasjournal}   

\end{document}